\documentclass[a4paper,12pt]{article}
\usepackage[english]{babel}
\usepackage{url}
\usepackage{layaureo}
\usepackage{fullpage}

\usepackage{graphicx}
\usepackage{color}
\usepackage{multirow}
\usepackage{amsthm}
\usepackage{amsmath}
\usepackage{amsfonts}
\usepackage{amssymb}
\usepackage{graphicx}
\usepackage{setspace}
\usepackage{natbib}
\usepackage{booktabs}
\usepackage[colorlinks,breaklinks]{hyperref}
\hypersetup{
    colorlinks,%
    citecolor=black,%
    filecolor=black,%
    linkcolor=black,%
    urlcolor=black
}
\usepackage{lscape}
\usepackage{ctable}

\numberwithin{equation}{section}

\newcommand{\R}{\mathbb{R}}

\allowdisplaybreaks

\begin{document}

\begin{center}
{\noindent \textbf{\Large Multivariate non-Gaussian models\\  for financial applications}}




\vspace{18pt}

{\noindent Michele Leonardo Bianchi\textsuperscript{a,*}, Asmerilda Hitaj\textsuperscript{b}, 
 Gian Luca Tassinari\textsuperscript{c}}\\
\vspace{10pt}
{\noindent \small\textsuperscript{a}\it Regulation and Macroprudential Analysis Directorate,}
{\it Bank of Italy, }\\
{\it micheleleonardo.bianchi@bancaditalia.it}\\
{\noindent \small\textsuperscript{*} Corresponding author}\\
\vspace{10pt}
{\noindent \small\textsuperscript{b}\it Department of Statistics and Quantitative Methods,}
{\it University of Milano-Bicocca, }\\
{\it asmerilda.hitaj1@unimib.it}\\
%
\vspace{10pt}
{\noindent \small\textsuperscript{c}\it Department of Economics, Department of Statistical Sciences ``Paolo Fortunati'',\\ and Department of Management, University of Bologna,}\\
{\it gianluca.tassinari2@unibo.it}\\
\vspace{10pt}
This version: \today\\
\vspace{10pt}
\end{center}

\noindent \textbf{Abstract.} In this paper we consider several continuous-time multivariate non-Gaussian models applied to finance and proposed in the literature in the last years. We study the models focusing on the parsimony of the number of parameters, the properties of the dependence structure, and the computational tractability. For each model we analyze the main features, we provide the characteristic function, the marginal moments up to order four, the covariances and the correlations. Thus, we describe how to calibrate them on the time-series of log-returns with a view toward practical applications and possible numerical issues. To empirically compare these models, we conduct an analysis on a five-dimensional series of stock index log-returns.

\vspace{25pt}


\newpage
\section{Introduction}

Many problems of practical interest in finance, such as  portfolio selection, multi-asset derivative pricing, or the estimation of systemic risk measures are multi-dimensional. The multivariate normal model is usually applied to solve these problems, mainly because both the theoretical and practical complexity of a model increases as soon as one moves from a Gaussian to a non-Gaussian framework. The multivariate normal distribution has two main drawbacks: (1) its margins are normally distributed, therefore it is not capable of capturing heavy tails and asymmetries (negative skewness); (2) its dependence structure is symmetric, it is not capable of capturing asymmetry of dependence during extreme market movements and the dependence of tail events. \cite{bedendo2010pricing} showed that in calm market conditions the choice of the dependence structure (i.e. the copula function) does not significantly affect option prices. Conversely, in volatile market scenarios both linear correlation and tail dependence strengthen and the specification of the dependence structure becomes much more relevant. For example, the dependence between the components of a portfolio and the components of a basket have an impact on portfolio risk measurement, in derivative pricing, or in the systemic risk estimation. Thus, multivariate distributions able to capture also non-linear dependence among margins are necessary. In light of the above, in this paper we analyze different multivariate non-Gaussian model based on L\'evy processes allowing for heavy tails and asymmetries of both margins and dependence. For all analyzed models we look for consistent and computationally efficient estimation procedures.

Notably for a multivariate model desirable properties are: (a) the density function can be written in closed form or in quasi-closed form (i.e. evaluated through an efficient and well-known numerical method); (b) the characteristic function has a closed form allowing one to explore the properties of the model or the change of measure needed to price derivative contracts; (c) the univariate distribution of the portfolio defined as weighted sum of the margins can be easily computed and applied to evaluate portfolio risk measures and to solve portfolio optimization problems; (d) the model is able to explain four stylized facts about financial time-series, that is heavy tails, negative skewness, asymmetric dependence, and volatility clustering (see \cite{allen2014four} and \cite{btf2015riding}); (e) the model can be extended to price derivatives or, at least, there is an efficient method to draw random samples from the model; (f) the number of parameters with respect to the multivariate normal does not increase too much (possibly linearly) by the number of margins, that is the model should be flexible enough but it should not be overparameterized; (g) there is at least one robust estimation algorithm and, ideally, there exists a package written in some commonly used programming language allowing one to perform the estimation.  

It should be noted that in this paper we review only continuous-time models based on L\'evy processes that usually are not capable of capturing neither the volatility clustering effect nor the leverage effect. The volatility clustering effect is the tendency of large changes in asset prices (either positive or negative) to be followed by large changes, and small changes to be followed by small changes. The leverage effect is the empirically observed fact that negative shocks have a stronger impact on the variance than positive shocks of the same magnitude (i.e. bad news raises the future volatility more than good news). From a practical perspective, it is possible to add volatility clustering dynamics by preliminary filtering the log-return data through a GARCH model and then by calibrating the multivariate models described in this paper on the standardized residuals, as done in \cite{btf2015riding}.

A multivariate non-Gaussian model can be built by following different approaches. For example, it is possible (1) to  consider a linear combination of independent L\'evy processes, as done for example by \cite{kawai2009multivariate}, \cite{kaishev2013levy}, \cite{ballotta2016multivariate} and \cite{loregian2018estimation}; (2) to time-change a multivariate L\'evy process with a univariate or a multivariate subordinator (e.g. \cite{barndorff2001multivariate}, \cite{luciano2010multivariate}, \cite{hitaj2016multivariate}, and \cite{semeraro2019note}); (3) to build the multivariate process by defining its L\'evy measure, as done by \cite{rosinski2007tempering} and \cite{bianchi2011tempered}; (4) to specify the models for the margins separately from the copula that links the margins (i.e. the dependence structure) to form a multivariate model (see \cite{genest2009advent}, \cite{patton2012review} and \cite{tankov2016copula} for a review). In this paper we consider only the first two approaches. Even if the third approach is elegant from a theoretical perspective, it is not simple to use in practical applications. Furthermore, copula methods provide a simple framework for thinking about dependence but it is not always clear how to find a change of measure to price derivatives.  Furthermore, as observed by \cite{mikosch2006copulas} and at least in the application analyzed here, there is no particular advantage of using copulas when dealing with multivariate distributions or processes. 

In this work we discuss and empirically assess the fitting performance of several continuous-time multivariate heavy-tailed and semi heavy-tailed models applied to finance and proposed in the literature in the last years. For each model we analyze the main properties and the more useful formulas, that is: the characteristic function, the probability density function (in the case it can be written in closed form), the marginal moments up to order four, the covariance and the correlations.  After having described how to calibrate these models, analyzed the computational tractability and possible numerical issues, we empirically compare these models on a real market dataset on the basis on some fitting error measures. Just for comparison purposes, we calibrate also the multivariate normal model.

The paper is organized as follows. In Section \ref{sec:MeanVarianceMixture} we review the normal mean-variance mixture models, where the mixing variable has a semi-heavy tailed distribution, and in Section \ref{sec:GMeanVarianceMixture} we consider extensions based on a multivariate mixing distribution referred to as one factor subordinated models. Both models belong to the class of time-changed L\'evy  models: while the former has a quasi-closed formula for the joint density function, the latter does not but it may have a greater flexibility in fitting market data thanks to its richer dependence structure. A multivariate model based on tempered stable distribution and its major capability to fit the margins is discussed in Section \ref{multiv}. A further extension is proposed in Section \ref{sec:GVG} where a more flexible multifactorial model is considered. This model is able to capture dependence separately and independently both in positive and negative jumps and in their finite and infinite activity components. Multivariate models based on linear combinations of L\'evy processes are presented in Section \ref{linear}. Differences in estimation methods are described in Section \ref{estimMethod}. In Section \ref{empan} we describe the data analyzed in the empirical study, we discuss the main empirical results and we identify some computational issues. After having summarized the main results, Section \ref{sec:Conclusions} concludes. Finally, in the Appendix we provide the formulas of expected value, variance, skewness, excess kurtosis and correlation of the models described in the paper.

\section{Normal mean-variance mixture models}\label{sec:MeanVarianceMixture}

The multivariate non-normal distributions analyzed in this section are a generalization of the multivariate normal distribution known as multivariate normal mean-variance mixtures. These distributions share much of the structure of the multivariate normal distribution, but they allow asymmetry, heavy tails and both linear and non-linear dependence. In particular, a random vector $Y$ has a multivariate normal mean-variance mixture distribution (NMV) if the following equality in law holds
\begin{equation}\label{eq:NMV}
Y=\mu + \theta S + \sqrt{S}QZ,
\end{equation}
where $\mu$, $\theta$ $\in \R^n$, $Q$ is a square matrix of order $n$ such that $QQ'$ is positive definite, $S$ is a positive random variable, $Z\sim N(0, I_n)$ and $S$ is independent from $Z$.
Furthermore, if the mixing variable $S$ is infinitely divisible then $Y$ is infinitely divisible and its law uniquely determines a time-changed L\'evy process whose subordinator at time one has the law of $S$ (see \cite{barndorff2001multivariate}). Therefore, in order to model asset returns it is possible to build an $n$-dimensional L\'evy process whose increments follow an infinitely divisible NMV distribution by simply time-changing a multivariate Brownian motion with a common $one$-dimensional subordinator. While in \cite{luciano2006multivariate} and \cite{tassinari2013qf} a model with independent Brownian motions was proposed, in \cite{leoni2008multivariate}, \cite{tassinari2009thesis}, \cite{wu2009closed}, \cite{tassinari2014maf}, \cite{tb2014ijtaf}, \cite{btf2015riding}, and \cite{bianchi2017forward} correlated Brownian motions were considered. 
Furthermore, according to \cite{frahm2004generalized}, this family of distributions belongs to the class of elliptical variance-mean mixtures. Elliptical and generalized elliptical heavy-tailed distributions have been widely studied (see e.g. \cite{kring2009multi}, \cite{dominicy2013inference}, \cite{bstff2019handbook}).
Let $Y = \{Y_t, t\geq 0\}$ be a multivariate process such that the following equality holds  
\begin{equation}\label{eq:OneTimeChange}
\begin{split}
Y_t &\;=\mu t + \theta S_{t} + D_{\sigma} W_{S_{t}},\\
\end{split}
\end{equation}
where $S = \{S_t, t\geq 0\}$ is a $one$-dimensional subordinator, $W = \{W_t, t\geq 0\}$ is an $n$-dimensional Wiener process with $corr\left[{W_{j,t},W_{k,t}}\right]=\rho_{jk}$ independent from $S$, and $D_{\sigma}$ is a diagonal matrix with diagonal elements $\sigma_j>0$ for $j=1,...,n$.
For each discrete time step $\Delta t$ the distribution of the increments of the process belongs to the NMV family
\begin{equation*}
Y_{\Delta t} = \mu \Delta t + \theta S_{\Delta t} + \sqrt{S_{\Delta t}} D_{\sigma} A Z,	
\end{equation*}
where $S_{\Delta t}$ denotes the distribution of the subordinator increments which is independent from $Z$, $A$ is the lower Cholesky decomposition of a correlation matrix $\Omega$, that is, $\Omega^{1/2} = A$ and $Q = D_{\sigma}A=\Sigma^{1/2}$.
The characteristic function of $Y_t$ defined in equation (\ref{eq:OneTimeChange}) is given by
\begin{equation}\label{eq:Chf}
 \Psi_{Y_t}\left(u\right)=\exp\left(itu'\mu + tl_{S_1}(\varphi(u))\right),
\end{equation}
where $l_{S_1}\left(.\right)$ is the Laplace exponent of the subordinator, and $\varphi\left(u\right)$ is the characteristic exponent of the multivariate Brownian motion, that is
\begin{equation}\label{eq:ChExpMBM1}
\begin{split}
g\left(u\right)&\;= i u'\theta-\frac{1}{2} u'\Sigma u\\  
&\;= \sum_{j=1}^n i u_j\theta_{j}-\frac{1}{2}\sum_{j=1}^n \sum_{k=1}^n u_{j} u_k \sigma_{j}\sigma_{k}\rho_{jk},
\end{split}
\end{equation}
where $u \in {\R}^n$ and the matrix $\Sigma$ has elements $\Sigma_{jk} = \sigma_{j}\sigma_{k}\rho_{jk}$. Since $\Sigma$ is a variance-covariance matrix, we can rewrite equation (\ref{eq:ChExpMBM1}) using matrix notation, as follows
\begin{equation*}
\varphi\left(u\right)= i u'\theta-\frac{1}{2} u'D_{\sigma}\Omega D_{\sigma} u,
\end{equation*}
where $D_{\sigma}$ is a diagonal matrix with diagonal $\sigma\in\R_{+}^n$, and $\Omega$ is the correlation matrix of the Brownian motions with elements $\rho_{jk}$.

\subsection{The multivariate generalized hyperbolic distribution}

The generalized hyperbolic (GH) distribution, introduced by \cite{barndorff1977exponential}, has received a lot of attention in the financial-modeling literature (see \cite{eberlein1995hyperbolic}, \cite{prause1999generalized}, and \cite{eberlein2002generalized}). Many well known distributions, like for example the student's $t$, the skew-$t$, the variance gamma (VG) and the normal inverse Gaussian (NIG), belong to the GH parametric family. In this section we review the multivariate extension of the GH distribution.

Let $S=\{S_t,\ t\geq0\}$ be a generalized inverse Gaussian process (GIG), i.e., a L\'evy process in which the law of $S_1$ is generalized inverse Gaussian with parameters $\epsilon$, $\psi$, $\chi$, where $\psi$ and $\chi$ are both nonnegative and not simultaneously 0. We denote the law of $S_1$ as $GIG\left(\epsilon, \chi, \psi \right)$. The density function of $S_1$ is
\begin{equation*}
f(x;\epsilon, \psi, \chi)= \frac{1}{2K_{\epsilon}\left(\sqrt{\chi\psi}\right)}\left(\frac{\psi}{\chi}\right)^{\frac{\epsilon}{2}} x^{\epsilon-1}\exp\left(-\frac{1}{2}\left(\frac{\chi}{x} + \psi x\right)\right), x>0,  
\end{equation*}
and its characteristic function is
\begin{equation}\label{eq:CfGIG}
\Psi_{S_1}{(u)}=\left(1-\frac{2iu}{\psi}\right)^{-\frac{\epsilon}{2}}\frac{K_{\epsilon}\left(\sqrt{\chi(\psi-2i u)}\right)}{K_{\epsilon}\left(\sqrt{\chi\psi}\right)}.
\end{equation}

If in equation (\ref{eq:OneTimeChange}) we select a subordinator $S = \{S_t, t\geq 0\}$ such that the characteristic function of $S_1$ is (\ref{eq:CfGIG}), then the process $Y = \{Y_t, t\geq 0\}$ is referred to as the multivariate generalized hyperbolic (MGH) process with parameters ($\epsilon$, $\chi$, $\psi$, $\theta$, $\mu$, $\Sigma$). 

Finally, using (\ref{eq:Chf}) we get the characteristic function of the MGH process with linear drift 
\begin{equation}\label{eq:ChfMGH}
\begin{split} 
\Psi_{Y_t}\left(u\right)&=\exp\left(i u' \mu t\right)\left(1-\frac{2}{\psi}\left(iu'\theta-\frac{1}{2} u'\Sigma u\right)\right)^{-\frac{\epsilon t}{2}} \left( \frac{K_{\epsilon}\left(\sqrt{\chi\left(\psi-2\left(iu'\theta-\frac{1}{2} u'\Sigma u\right)\right)}\right)}{K_{\epsilon}\left(\sqrt{\chi\psi} \right)}\right) ^t. 
\end{split} 
\end{equation}  
Setting $u_i=0$, $\forall i\neq j$, into (\ref{eq:ChfMGH}) we get the characteristic function of the log-return process of the $j$-th underlying asset
\begin{equation}\label{eq:CfMGHMar} 
\Psi_{Y_{j,t}}\left(u_j\right)=\exp\left(i u_j \mu_j t\right)\left(1-\frac{2}{\psi}\left(iu_j\theta_j-\frac{1}{2} u_j^2\sigma_j^2\right)\right)^{-\frac{\epsilon t}{2}} \left( \frac{ K_{\epsilon} \left(\sqrt{\chi\left(\psi-2\left(iu_j\theta_j-\frac{1}{2}u_j^2\sigma_j^2\right)\right)}\right)}{K_{\epsilon}\left(\sqrt{\chi\psi} \right)}\right)^t.
\end{equation}
Setting $t=1$ into (\ref{eq:ChfMGH}) and into (\ref{eq:CfMGHMar}) we get the characteristic function of the MGH and GH distributions. 

Comparing the characteristic function of the MGH distribution with the one of $Y_t$ we can notice that the GH distribution is infinitely divisible but not closed under convolution. Thus, if $Y_1$ is a MGH random variable, $Y_t$ is not. If $\epsilon=-1/2$, $G_1$ follows an inverse Gaussian distribution with parameters $\gamma=\sqrt{\chi}$ and $\eta=\sqrt{\psi}$. If $\chi=0$, $G_1$ follows a gamma distribution $\alpha=\epsilon$ and $\beta=\psi/2$. In the first case we get the multivariate normal inverse Gaussian (MNIG) model and in the second one the multivariate variance gamma (MVG) considered in \cite{tb2014ijtaf}.\\

\subsection{The multivariate normal tempered stable distribution}

The tempered stable family was introduced by \citeauthor{boyarchenko2000truncated}  (\citeyear{boyarchenko2000truncated}, \citeyear{boyarchenko2002nongaussian}) and studied in deep by \cite{rosinski2007tempering}. The multivariate model discussed in this section is based on the classical tempered stable (CTS) distribution as described by \cite{kim2012measuring}, \cite{btf2015riding} and \cite{bianchi2017forward} and extends the works of \cite{prause1999generalized}, \cite{leoni2008multivariate} and \cite{wu2009closed} to the CTS case. We refer to this model as multivariate normal tempered stable (MNTS). We observe that the elliptical tempered stable (ETS) distribution defined in \cite{fallahgoul2016elliptical} is a subclass of the multivariate symmetric normal tempered stable (MSNTS) distribution, and a symmetric MNTS is a subclass of the tempered infinitely divisible family introduced by \cite{bianchi2011tempered}.

The process $S = \{S_t, t\geq 0\}$ is said to be a CTS subordinator with parameters $\alpha$, $\lambda>0$, $C>0$,  $0<\alpha<1$ if the characteristic function of $S_t$ is given by
\begin{equation}\label{eq:ChfCTSplusSubordinator}
\phi_{S_t}(u) = E[\exp(iuS_t)] = \exp\left( tC\Gamma(-\alpha)((\lambda - iu)^{\alpha} -
\lambda^{\alpha}))\right),
\end{equation}
where $u\in\R$ and $\Gamma(\cdot)$ is the gamma function. From equation (\ref{eq:ChfCTSplusSubordinator}) it is possible to compute the Laplace exponent of the CTS subordinator
\begin{equation*} 
l_{S_t}(u)=\ln\phi_{S_t}{\left(-iu\right)}=tC\Gamma(-\alpha)((\lambda - u)^{\alpha} -\lambda^{\alpha}).
\end{equation*}

If in equation (\ref{eq:OneTimeChange}) we select a subordinator $S = \{S_t, t\geq 0\}$ with characteristic function (\ref{eq:ChfCTSplusSubordinator}), then the process $Y = \{Y_t, t\geq 0\}$ is referred to as the MNTS process with parameters ($a$, $\lambda$, $C$, $\theta$, $\mu$, $\Sigma$). If one selects $\alpha=a/2$ and $C=\lambda^{1-\alpha}/\Gamma(1-\alpha)$, equation (\ref{eq:ChfCTSplusSubordinator}) can be written as
\begin{equation*}
\phi_{S_t}(u) = \exp\left(- t\frac{2\lambda^{1-\frac{a}{2}}}{a}\left((\lambda - iu)^{\frac{a}{2}} - \lambda^{\frac{a}{2}}\right)\right),
\end{equation*}
and the multivariate distribution $Y_{\Delta t}$ defined as  
\begin{equation*}
Y_{\Delta t}  = \mu {\Delta t}  + \theta(S_{\Delta t}  - {\Delta t} ) + \sqrt{S_{\Delta t}} D_{\sigma} A Z,	
\end{equation*}
is the MNTS distribution analyzed by \cite{kim2012measuring}.

Using (\ref{eq:Chf}) we get the characteristic function of the MNTS process with linear drift
\begin{equation}\label{eq:ChfMNTS}
\Psi_{Y_t}\left(u\right)=\exp\left(t \left( iu'\mu + C\Gamma\left(-{\frac{a}{2}}\right)\left(\left(\lambda - iu'\theta + \frac{1}{2} u'\Sigma u\right)^{\frac{a}{2}} - \lambda^{\frac{a}{2}} \right)\right)\right).
\end{equation}
Setting $u_i=0$, $\forall i\neq j$, into (\ref{eq:ChfMNTS}) we get the characteristic function of the $j$-th marginal distribution
\begin{equation}\label{eq:CfMntsMar} 
\Psi_{Y_{j,t}}\left(u_j\right)=\exp\left(t \left( iu_j\mu_j + C\Gamma\left(-{\frac{a}{2}}\right)\left(\left(\lambda - iu_j\theta_j + \frac{1}{2} u_j^2\sigma_j^2\right)^{\frac{a}{2}} - \lambda^{\frac{a}{2}} \right)\right)\right).
\end{equation}

If $\omega=1/2$, $S_1$ follows an inverse Gaussian distribution with parameters $\gamma=-\frac{C\Gamma(-\omega)}{\sqrt{2}}$ and $\eta=\sqrt{2 \lambda}$, and $Y_1$ follows the MNIG distribution with parameters $\alpha=-\frac{C\Gamma(-\omega)}{\sqrt{2}}$ and $\beta=\sqrt{2 \lambda}$, and $Y_1$ follows the MVG distribution described in \cite{tb2014ijtaf}.

\subsection{Main properties and estimation methods}

It is important to note that unlike the MNTS case, if the increments of the process at a given time scale (e.g. daily) follow a MGH distribution, on a different time scale (e.g. yearly) the increments follow an infinitely divisible distribution different from the MGH (see \cite{cont2003financial}), making the MGH distribution less convenient when one needs to work with data with different time scales. This problem can arise, for example, in options pricing models that make use of both daily returns and implied volatilities (see \cite{tb2014ijtaf} and \cite{bianchi2017forward}).

As shown in \cite{embrechts2005quantitative} and \cite{kim2012measuring}, the portfolio constructed as a linear combination of GH (NTS) margins has a GH (NTS) distribution. Thus, this model can be easily applied to evaluate widely known portfolio risk measures and to solve asset allocation and portfolio optimization problems (see \cite{bianchi2017forward}).

Unlike the MGH random variable, it is not possible to obtain in closed form the probability density function of the MNTS random variable. However, the density of a MNTS random variable can be obtained by a numerical integration that combines the density of a multivariate normal distribution and the density of a univariate tempered stable mixing distribution, which can be evaluated by means of a fast Fourier transform (FFT) (see \cite{stoyanov2004numerical} and \cite{bianchi2013tsou}). 

In estimating these models, it is usually not possible to resort to direct maximization of the likelihood function as the number of parameters is large. To overcome this obstacle, in the estimation of the parameters of the multivariate normal mean-variance mixture distributions, the use of the expectation-maximization (EM) maximum likelihood estimation method is particularly convenient as it allows to find the parameters of the multivariate Gaussian distribution and those of the mixing distribution separately (see \cite{protassov2004based}, \cite{hu2005calibration}, and \cite{embrechts2005quantitative} in the MGH case, and  \cite{btf2015riding} in the MNTS). 

In the elliptical TS case \cite{fallahgoul2016elliptical} estimated $\mu$ and $\Sigma$  using the sample mean vector and sample variance-covariance matrix for stock market returns. The parameters $\alpha$ and $\lambda$ were obtained by considering the average of the margin estimates.  \cite{kim2012measuring} conducted an empirical analysis on the Dow Jones Industrial Average (DJIA) index and  29 of the 30 component stocks. They estimated the parameters $\alpha$ and $\lambda$ on the DJIA index returns and the vector $\theta$ was estimated on the margins. Finally, both $\mu$ and $\Sigma$ were estimated by considering the sample covariances together with the univariate estimates. However, we do not explore this method in our empirical study, since we will rely on the EM estimation approach as it will be described in Section \ref{sec:EM}.

\section{Generalized NMV mixture models}\label{sec:GMeanVarianceMixture}

A multivariate NMV distribution is based on a common $one$-dimensional mixing variable.  This corresponds to a multidimensional return process with a unique stochastic time-change, which implies the uniqueness of the business time for all assets. As shown by \cite{harris1986cross} this feature seems to be inconsistent with empirical evidence. Starting from the work of \cite{semeraro2010multivariate} in which a generalization of the VG process is discussed, \cite{luciano2010generalized} proposed a generalization of the definition of NMV distribution based on an $n$-dimensional mixing variable. Recently, \cite{rathgeber2019financial} conducted a large simulation study on these models in order to identify the best fitting method for multivariate models.

A random vector $Y$ has a multivariate generalized normal mean-variance mixture distribution (GNMV) if the following equality in law holds 
\begin{equation}\label{eq:GNMV}
Y=\mu + M D_G \theta + QD_{\sqrt{G}}Z,
\end{equation}
with $\mu$, $\theta$ $\in \R^n$, $M$ and $Q$ are square matrices of order $n$, $QQ'$ is positive definite, $G$ is an $n$ dimensional positive random vector whose $j$-th component is $G_j$, $D_G$ and $D_{\sqrt{G}}$ are diagonal matrices with diagonal elements $G_j$ and $\sqrt{G_j}$, respectively, and $Z\sim N(0, I_n)$ is independent from $G$. If the mixing variable $G$ is infinitely divisible then $Y$ is infinitely divisible and its law uniquely determines a L\'evy process. 

Following \cite{luciano2010generalized} the characteristic function of the random variable $Y$ can be written as  
\begin{equation*}
\Psi_{Y}\left(u\right)=\exp\left(iu'\mu\right) \exp\left(l_{G}\left(i D_{\theta}M'u-\frac{1}{2}D_{Q'u} {Q'u}\right)\right),
\end{equation*}
where $l_{G} \left(.\right)$ is the Laplace exponent of the multivariate mixing variable $G$, $D_{\theta}$ and $D_{Q'u}$ are diagonal matrices with diagonal elements the vectors $\theta$ and $Q'u$, respectively. If we set $M=I_n$ and $G_j=S$ for all $j$ in (\ref{eq:GNMV}) we obtain (\ref{eq:NMV}). 

\cite{barndorff2001multivariate} proved that a random vector $Y$ has GNMV distribution if and only if it is the law at time one of a L\'evy process obtained by subordination of a $\R_{+}^n$-parameter Brownian motion with a multidimensional subordinator whose distribution is given by $G$. \cite{luciano2010generalized} and \cite{luciano2010multivariate} built multivariate L\'evy processes with GH, compound Poisson, NIG and VG margins using the multivariate subordination technique. In particular, they proposed two different techniques to build $n$-dimensional L\'evy processes through subordination leading to different multivariate models with the same marginal processes. We refer to these two class of 
processes as the $\alpha$-models and the $\rho\alpha$-models. Recently, an extension of the $\alpha$-model based on the VG distributional assumption and weak-subordination and allowing a wider range of dependence has been applied to finance by \cite{michaelsen2018marginal}, \cite{madan2018instantaneous} and \cite{buchmann2018calibration}.

\subsection{The \texorpdfstring{$\alpha$}{}GH distribution}\label{sec:alphaMeanVarianceMixture}

\cite{lo2000trading} provided empirical evidence that business time as measured by trades presents a significant common component. \cite{semeraro2010multivariate}, \cite{luciano2010generalized} and \cite{luciano2010multivariate} proposed to build multivariate subordinators able to capture  both a time-change common to all assets and an idiosyncratic one. In particular, they used the random additive effect distributions proposed by \cite{barndorff2001multivariate} to get a multivariate stochastic clock $G = \{G_t, t\geq 0\}$ containing both a common and an asset specific time-change:
\begin{equation}\label{eq:addmultsub}
G_ {t}=X_{t}+\alpha S_t, 
\end{equation}
where $X = \{X_t, t\geq 0\}$ is an $n$-dimensional subordinator with independent components, $S = \{S_t, t\geq 0\}$ is a $one$-dimensional subordinator independent by $X$, $\alpha$ is a $n\times 1$ vector with positive elements. Let $Y = \{Y_t, t\geq 0\}$ be a multivariate process such that the following equalities hold  
\begin{equation}\label{eq:FactorBasedIND}
\begin{split}
Y_t &\;=\mu t + Y_{t}^{I}\\ 
&\;= \mu t + B_{G_{t}}\\
&\;= \mu t + D_{G_{t}} \theta+D_{\sigma} W_{G_{t}},
\end{split}
\end{equation}
where
\begin{itemize}
\item $Y^I = \{Y_t^I, t\geq 0\}$ is constructed by subordinating an $n$-dimensional arithmetic Brownian motion $B=\{B_t, t\geq 0\}$, where $B_t=\theta t+D_{\sigma}W_t$, with independent components with the subordinator (\ref{eq:addmultsub});
\item $W = \{W_t, t\geq 0\}$ is an $n$-dimensional Wiener process with $corr\left[ {W_{j,t},W_{k,t}}\right]=0$  for $j\neq k$;
\item $D_{\sigma}$ is a diagonal matrix with diagonal elements $\sigma_j\in\R_{+}$ for all $j$,
\item $\theta\in \R^n$ is a vector of parameters. 
\end{itemize} 

For each discrete time step $\Delta t$ the distribution of the increments of the process is given by
\begin{equation*}
Y_{\Delta t}= \mu\Delta t + D_{G_{\Delta t}}\theta+D_{\sigma} D_{\sqrt{G_{\Delta t}}} Z,\\
\end{equation*}
where $G_{\Delta t}$ denotes the distribution of the subordinator increments which is independent of $Z$. The distribution of $Y_{\Delta t}$ belongs to the class of the GNMV distribution with $M=I_n$ and $Q=D_{\sigma}$.

The characteristic function of $Y_t$ defined in equation (\ref{eq:FactorBasedIND}) is given by
\begin{equation}\label{eq:ChfFactorBasedIND}
\begin{split}
 \Psi_{Y_t}\left(u\right)&\;=\exp\left(itu'\mu\right)\Psi_{Y_t^I}\left(u\right)\\
 &\;= \exp\left(itu'\mu\right)\exp\left(t\sum_{j=1}^n l_{X_{j,1}}(\psi_j(u_j))\right)\exp\left( tl_{S_1}(\sum_{j=1}^n \alpha_j \psi_j(u_j))\right)\\
 &\;= \exp\left(itu'\mu\right)\exp\left(t\sum_{j=1}^n l_{X_{j,1}}(iu_j\theta_j-\frac{1}{2}u_j^2\sigma_j^2)\right)\exp\left( tl_{S_1}(\sum_{j=1}^n \alpha_j (iu_j\theta_j-\frac{1}{2}u_j^2\sigma_j^2))\right),
 \end{split}
\end{equation}
where $l_{X_{j,1}}\left(.\right)$ and $l_{S_1}\left(.\right)$ are the Laplace exponents of the subordinators $X_{j,t}$ and $S_t$, respectively.
Choosing $X_{j,1}$, $S_1$ and $\alpha_j$ opportunely, \cite{luciano2010generalized} and \cite{luciano2010multivariate} proposed different multivariate models with GH, compound Poisson, NIG and VG  margins. We review only the GH model, and we refer to it as $\alpha$GH model. Following \cite{luciano2010generalized} we build a multivariate subordinator $G = \{G_t, t\geq 0\}$ on $\R_+^n$ with dependent GIG margins $G_j = \{G_{j,t}, t\geq 0\}$, $j=1,...,n$, by defining
\begin{equation}\label{eq:MGIG} 
G_{j,t} =X_{j,t}+\alpha_j S_{t}=R_{j,t}+P_{j,t}+\frac{1}{\psi_j}S_{t},
\end{equation}
where $R_{j,1}$, $P_{j,1}$ and $S_{1}$ are independent with $R_{j,1}\sim GIG \left(-\epsilon, \chi_j, \psi_j\right) $, $P_{j,1}\sim\Gamma\left(\epsilon-a,\frac{\psi_j}{2}\right)$, and $S_{1} \sim\Gamma\left(a,\frac{1}{2}\right)$. 
If $\epsilon > 0$, $\psi_j > 0$, $0 < a < \epsilon$, and $\chi_j \geq 0$ for all $j$, then $G_{j,1}\sim GIG \left(\epsilon, \chi_j,\psi_j\right)$, that is all the margins of the multivariate subordinator at time one follow a generalized inverse Gaussian law. \\Using (\ref{eq:ChfFactorBasedIND}) and setting $t=1$ we get the characteristic function of the $\alpha$GH distribution
\begin{equation}\label{eq:CfalphaIGH} 
\begin{split}
\Psi_{Y_{1}}\left(u\right)&\;= \prod_{j=1}^n \left(1 - \frac{2}{\psi_j}\left( iu_j\theta_j - \frac{1}{2}\sigma_j^2 u_j^2\right) \right)^{a-\frac{\epsilon}{2}}\frac{ K_{\epsilon} \left(\sqrt{\chi_j\left(\psi_j-2\left(iu_j\theta_j-\frac{1}{2}u_j^2\sigma_j^2\right)\right)}\right)}{K_{\epsilon}\left(\sqrt{\chi_j\psi_j} \right)}\\  &\;\quad \exp\left(iu'\mu\right) \left(1 -   \sum_{j=1}^n \frac{2}{\psi_j}\left( i u_j\theta_j-\frac{1}{2}\sigma_j^2 u_j^2\right)\right)^{-a}.
\end{split}
\end{equation}
Setting $u_i=0$, $\forall i\neq j$, into (\ref{eq:CfalphaIGH}) we get the characteristic function (\ref{eq:CfMGHMar}) of the GH law.  

Setting $\epsilon=1$ the marginal processes are hyperbolic and we get the $\alpha$HYP. 
If $a\rightarrow 0$ the $\alpha$GH process degenerates into the MGH model with independent univariate GH processes.
If $\chi_j\rightarrow 0$ for all $j$, the $\alpha$GH process degenerates into the $\alpha$VG.
If one sets $\chi_j=\delta_j^2$, $\psi_j=\alpha_j^2-\beta_j^2$, $\theta_j=\beta_j$, $\mu_j =0$, $\sigma_j = 1$ for all $j$, we get the $\alpha$GH process of \cite{luciano2010generalized}. 

The estimation of this model in \cite{luciano2010generalized} was performed in two steps by fixing $\epsilon=1$. The restrictions on the parameters of the random variables $R_{j,1}$, $P_{j,1}$ and $S_{1}$ in the right hand side of \eqref{eq:MGIG} ensure that the margins are still GH distributed. \cite{Guillaume2013}, following \cite{luciano2010multivariate} and \cite{semeraro2010multivariate},  but removing the restrictions on single variable parameters, proposed the generalized $\alpha$VG model whose margins are no longer VG distributed but still result to be infinitely divisible. The same principle can be followed to generalize the $\alpha$NIG and the $\alpha$GH models.

\subsection{The \texorpdfstring{$\rho\alpha$}{}GH distribution}\label{sec:alpharhoMeanVarianceMixture}

The $\alpha$-models are obtained time changing a multivariate Brownian motion with independent components with the subordinator (\ref{eq:addmultsub}). The only source of dependence among different assets is due to the timing of the jumps. The $\rho\alpha$-models extend the $\alpha$-models allowing the dependence of both time and size of the jumps.

Let $Y = \{Y_t, t\geq 0\}$ be a multivariate process such that the following equalities hold 
\begin{equation*}
\begin{split}
Y_t &\;=\mu t + Y_{t}^{I} + Y_{t}^{\rho}\\ 
&\;= \mu t + B_{X_{t}}  + B_{S_t}^{\rho}\\
&\;= \mu t + D_{X_{t}} \theta+D_{\sigma} W_{X_{t}}+D_{S_{t}} \theta^{\alpha}+D_{\sigma^ {\alpha}} W_{S_t}^{\rho},\\
\end{split}
\end{equation*}
where
\begin{itemize}
\item $Y^I = \{Y_t^I, t\geq 0\}$ is constructed by subordinating an $n$-dimensional arithmetic Brownian motion $B=\{B_t, t\geq 0\}$ with an $n$-dimensional subordinator $X=\{X_t, t\geq 0\}$ with independent components $X_j=\{X_{j,t}, t\geq 0\}$;
\item $Y^{\rho} = \{Y_t^{\rho}, t\geq 0\}$ is constructed by subordinating an $n$-dimensional Brownian motion $B^{\rho}=\{B^{\rho}_t, t\geq 0\}$ with the common $one$-dimensional subordinator $S=\{S_t, t\geq 0\}$;
\item $W = \{W_t, t\geq 0\}$ and $W^{\rho}  = \{W^{\rho}_t, t\geq 0\}$ are independent $n$-dimensional Wiener processes, with $corr\left[ {W_{j,t},W_{k,t}}\right] =0$ and $corr\left[ {W_{j,t}^{\rho}, W_{k,t}^{\rho}}\right]  =\rho_{jk}$ for $j\neq k$;
\item $X=\{X_t, t\geq 0\}$ and $S=\{S_t, t\geq 0\}$ are an $n$-dimensional and a $one$-dimensional independent subordinators, independent of $W = \{W_t, t\geq 0\}$ and $W^{\rho}  = \{W^{\rho}_t, t\geq 0\}$;
\item $\theta^{\alpha}$, $\sigma^{\alpha}$ and $\alpha$ are $n$-dimensional vectors with $\theta^{\alpha} = \theta\times\alpha$, $\sigma^{\alpha}=\sigma\times\sqrt{\alpha}$ (the symbol $\times$ stands for the component-wise product of two vectors), where $\alpha_j\in\R_{+}$, for all $j$;
\item $D_{X_{t}}$, $D_{\sigma}$, $D_{S_{t}}$ and  $D_{\sigma^ {\alpha}}$ are diagonal matrices with diagonal elements $X_{j,t}$, $\sigma_j$, $S_{t}$, and $\sigma_j\sqrt{\alpha_j}$ respectively, for all $j$.
\end{itemize} 
The parameter $\alpha_j$ must be chosen so that 
\begin{equation}\label{eq:FactorBasedA}
\begin{split}
Y_{j,t} &\;=\mu_j t + Y_{j,t}^{I} + Y_{j,t}^{\rho}\\ 
&\;= \mu_j t + B_{j,X_{j,t}}  + B_{j,S_t}^{\rho}\\
&\;= \mu_j t + \theta_jX_{j,t}+\sigma_jW_{j,X_{j,t}}+\theta_j\alpha_j S_t+ \sigma_j \sqrt{\alpha_j} W_{j,S_t}^\rho\\
\end{split}
\end{equation}
can be written as a time-changed Brownian motion 
\begin{equation*}
Y_{j,t} = \mu_j t + \theta_j G_{j,t} + \sigma_j W_{G_{j,t}},	
\end{equation*}
where $G_{j,t} = X_{j,t} + \alpha_j S_{t}$ for all $j$.
For each discrete time step $\Delta t$ the distribution of the increments of the process can be written as 
\begin{equation*}
Y_{\Delta t}= \mu\Delta t + D_{X_{\Delta t}}\theta+D_{\sigma} D_{\sqrt{X_{\Delta t}}} Z_{(1)} +\theta^{\alpha} S_{\Delta t} + \sqrt{S_{\Delta t}} D_{\sigma}^{\alpha} A Z_{(2)},
\end{equation*}
where $X_{\Delta t}$ and $S_{\Delta t}$ denote the distributions of the subordinators increments, $Z_{(i)}$ ($i=1, 2$) are independent $N\left(0, I_n \right)$ random vectors, $A$ is the lower Cholesky decomposition of the correlation matrix of $W^\rho$. 

As shown in \cite{luciano2010multivariate}, the characteristic function of $Y_t$ defined in equation (\ref{eq:FactorBasedA}) is given by
\begin{equation}\label{eq:ChfFactorBased}
\begin{split}
 \Psi_{Y_t}\left(u\right)&\;=\exp\left(itu'\mu\right)\Psi_{Y_t^I}\left(u\right)\Psi_{Y_t^{\rho}}\left(u\right)\\
 &\;= \exp\left(itu'\mu\right)\exp\left(t\sum_{j=1}^n l_{X_{j,1}}(\varphi_j(u_j))\right)\exp\left( tl_{S_1}(\varphi^{\rho\alpha}(u))\right)\\
 &\;= \exp\left(itu'\mu\right)\exp\left(t\sum_{j=1}^n l_{X_{j,1}}(iu_j\theta_j-\frac{1}{2}u_j^2\sigma_j^2)\right)\exp\left( tl_{S_1}(iu'\theta^{\alpha}-\frac{1}{2}u'\Sigma^{\rho\alpha}u)\right),
 \end{split}
\end{equation}
where $l_{X_{j,1}}\left(.\right)$ and $l_{S_1}\left(.\right)$ are the Laplace exponents of the subordinators $X_{j,t}$ and $S_t$, respectively, and $\Sigma^{\rho\alpha}=Var\left(B_1^{\rho}\right)$ is a positive definite matrix with elements $\Sigma_{jk}^{\rho\alpha}=\sigma_j \sigma_k \sqrt{\alpha_j} \sqrt{\alpha_k} \rho_{jk}$.

Choosing $X_{j,1}$, $S_1$ and $\alpha_j$ opportunely, \cite{luciano2010multivariate} and \cite{luciano2016dependence} proposed different multivariate models with compound Poisson, VG, NIG, and GH margins. We review only the last model and we refer to it as $\rho\alpha$GH model. 

Considering the GIG subordinator defined in (\ref{eq:MGIG}) and using (\ref{eq:ChfFactorBased}) for $t=1$, we get the characteristic function of the $\rho\alpha$GH distribution
\begin{equation}\label{eq:CfalpharhoGH} 
\begin{split}
\Psi_{Y_{1}}\left(u\right)&\;= \prod_{j=1}^n \left(1 - \frac{2}{\psi_j}\left( iu_j\theta_j - \frac{1}{2}\sigma_j^2 u_j^2\right) \right)^{a-\frac{\epsilon}{2}}\frac{ K_{\epsilon} \left(\sqrt{\chi_j\left(\psi_j-2\left(iu_j\theta_j-\frac{1}{2}u_j^2\sigma_j^2\right)\right)}\right)}{K_{\epsilon}\left(\sqrt{\chi_j\psi_j} \right)}\\  &\;\quad \exp\left(iu'\mu\right) \left(1 - 2 \left(i u'\theta^{\alpha}-\frac{1}{2} u'\Sigma^{\rho\alpha} u\right)\right)^{-a}.
\end{split}
\end{equation}
Setting $u_i=0$, $\forall i\neq j$, into (\ref{eq:CfalpharhoGH}) we get the characteristic function (\ref{eq:CfMGHMar}) of the GH law.

Setting $\epsilon=1$ the marginal processes are hyperbolic and we get the $\rho\alpha$HYP. 
If $a\rightarrow 0$ the $\rho\alpha$GH process degenerates into the MGH model with independent univariate GH processes.
If $\rho_{jk}=0$ for all $j\neq k$ then we obtain the $\alpha$GH model.
If $\chi_j\rightarrow 0$ for all $j$, the $\rho\alpha$GH process degenerates into the $\rho\alpha$VG process which includes both the MVG and the $\alpha$VG. Observe that it is not possible to obtain the MGH model and, since by construction $\epsilon$ must be positive,  it is not possible to obtain multivariate models with NIG marginal processes.
If one sets $\epsilon=\lambda$, $\theta^{\alpha}=\mu^{\rho}$, $\chi_j=\delta_j^2$, $\psi_j=\gamma_j^2-\beta_j^2$, $\theta_j=\beta_j$, $\mu_j =0$, $\sigma_j = 1$ for all $j$, we get the $\rho\alpha$GH process of \cite{luciano2010multivariate}.

\section{The multivariate mixed TS distribution}\label{multiv}

The multivariate mixed tempered stable (MMixedTS) distribution has been proposed in \cite{hitaj2016multivariate} and applied to portfolio selection in \cite{HMR2017senitivity}. This multivariate model is built on the basis of the standardized classical tempered stable (stdCTS) distribution.
A process $Y=\left\lbrace Y_t, t\geq 0 \right\rbrace $ with values in $\mathbb{R}^n$ is called MMixedTS if for  each margin $j$ the following equality holds
\begin{equation*}
Y_{j,t}=\mu_{j}t+\beta_{j}V_{j,t}+\sqrt{V_{j,t}}X_{j,t},\label{comp:mixedts}
\end{equation*}
where $V_{j}=\left\lbrace V_{j,t}, t\geq 0\right\rbrace$ is the $j$-th component of the multivariate subordinator $V=\left\lbrace V_t, t\geq 0\right\rbrace$, defined as 
\begin{equation*}
V_{j,t}=G_{j,t}+a_{j}\Lambda_t,
\end{equation*} 
in which $G_{j,t}$ and $\Lambda_t$ are nonnegative infinitely divisible random variables with
$G_{j,t}$ and $\Lambda_t$ mutually independent, $a_{j} \geq 0$ and
\begin{equation*}
X_{j,t}|V_{j,t}\sim stdCTS\left(\alpha_{j},\lambda_{+,j}\sqrt{V_{j,t}},\lambda_{-,j}\sqrt{V_{j,t}}\right).
\label{condXV}
\end{equation*}
for $j$ from $1$ to $n$.

In particular, if for each $j$,  $G_{j,t}\sim\Gamma(c_{j}t,m_{j})$, $\Lambda_t\sim\Gamma(\bar{n}t,k),$ and 
$a_{j}=\frac{k}{m_{j}}$, then $V_{j,t}\sim\ \Gamma(\left(c_{j}+\bar{n}\right)t,m_{j})$ 
that guarantees infinite divisibility, necessary for the definition of multivariate MixedTS-$\Gamma$.

Using matrix notation the MMixedTS distribution  can be written as
\begin{equation*}
Y=\mu+D_{\beta} V+D_V^{\frac{1}{2}}X
\end{equation*}
where $\mathbf{\mu} \in \mathbb{R}^n$, $D_{\beta} \in \mathbb{R}^{n\times n}$ with $D_{\beta}=diag\left(\beta_{1}, \ldots \beta_{n}\right)$, $V \in\mathbb{R}^n$ is a random vector with positive elements, $D_V$ is a random matrix positive defined, such that $D_V=diag\left(V_{1}, \ldots V_{n}\right)$, and $X$ is a stdCTS random vector.

The characteristic function of the MMixedTS process is
\begin{eqnarray*}
\Psi_{Y_t}(u)&=&\exp\left(i\sum\limits _{j=1}^{n}u_{j}\mu_{j}t+ tl_{\Lambda_1}\left(\sum\limits _{j=1}^{n}\left(i a_{j}u_{j}\beta_{j}+a_{j}\varphi_{stdCTS}\left(u_j;\lambda_{+,j},\lambda_{-,j},\alpha_{j}\right)\right)\right)\right)\nonumber\\
 &&\prod\limits _{j=1}^{n}\exp\left(tl_{G_{j,1}}\left(iu_{j}\beta_{j}+\varphi_{stdCTS}\left(u_j;\lambda_{+,j},\lambda_{-,j},\alpha_{j}\right)\right)\right), 
\end{eqnarray*}
where the $\varphi_{stdCTS}\left(u;\alpha,\lambda_{+},\lambda_{-}\right)$ is the characteristic exponent of a stdCTS random variable defined as 
\begin{equation*}
\varphi_{stdCTS}\left(u;\ \lambda_{+},\ \lambda_{-}, \ \alpha \right)=\frac{\left(\lambda_{+}-iu\right)^{\alpha}-\lambda_{+}^{\alpha}+\left(\lambda_{-}+iu\right)^{\alpha}-\lambda_{-}^{\alpha}}{\alpha\left(\alpha-1\right)\left(\lambda_{+}^{\alpha-2}+\lambda_{-}^{\alpha-2}\right)}\ +\frac{iu\left(\lambda_{+}^{\alpha-1}-\lambda_{-}^{\alpha-1}\right)}{\left(\alpha-1\right)\left(\lambda_{+}^{\alpha-2}+\lambda_{-}^{\alpha-2}\right)}.
\end{equation*}

\section{Multifactorial subordinated models}\label{sec:GVG}

As described in \cite{luciano2010generalized} a multivariate Brownian motion can be subordinated by considering a single factor $G$ defined as an $n$-dimensional positive random vector. This model can be extended to a multifactorial model as proposed by \cite{marfe2012generalized}. Further extensions of this model have been proposed by \cite{marfe2012multivariate} and \cite{guillaume2016sato}.  \cite{marfe2012generalized} introduced a multidimensional pure jump model with generalized variance gamma (GVG) margins able to capture dependence separately and independently both in positive and negative jumps and in their finite and infinite activity components. 

A multivariate generalized gamma (MGG) process is the L\'evy process $\hat{G}  = \{\hat{G}_t, t\geq 0\}$ on $\R_{+}^{n}$, where each component $\hat{G}_j = \{\hat{G}_{j,t}, t\geq 0\}$, $j=1,...,n$, is defined as the linear combination of independent subordinators, that is
\begin{equation*}\label{eq:GeneralizedGamma}
\begin{split}
\hat{G}_{j,t} =G_{j,t}+q_j G_{c,t}+G_{j,N_{j,t}}^* +p_jG_{c,N_{c,t}}^*= G_{j,t}+q_j G_{c,t}+X_{j,t} +p_jX_{c,t}
\end{split}
\end{equation*}
with $G_{j,1}\sim\Gamma\left(\frac{1-k_j}{q_j}-c_1,\frac{1}{q_j}\right)$, $G_{c,1}\sim\Gamma\left(c_1,1\right)$, $G_{j,1}^*\sim\Gamma\left(1,\frac{1}{p_j}\right)$, $G_{c,1}^*\sim\Gamma\left(1,1\right)$, $N_{j,1}\sim Poiss \left(\frac{k_j}{p_j}-c_2\right)$, $N_{c,1}^*\sim Poiss\left(c_2\right)$, $X_{j,1}\sim CP\left(\frac{k_j}{p_j}-c_2, 1, \frac{1}{p_j}\right)$ and $X_{c,1}\sim CP\left(c_2, 1, 1\right)$ where $0<c_1<\min_j\frac{1-k_j}{q_j}$, $0<c_2<\min_j \frac{k_j}{p_j}$, and $CP(\lambda, \alpha, \beta)$ denotes the law at time one of a compound Poisson process with jump intensity $\lambda$ and jump size $\Gamma(\alpha, \beta)$. 

The construction in equation (\ref{eq:GeneralizedGamma}) allows to express each margin as linear combination of two common factors, $G_{c,t}$ and $X_{c,t}$, and two idiosyncratic factors, $G_{j,t}$ and $X_{j,t}$. Each marginal process can be decomposed into the sum of an infinite activity component, $G_{j,t}+q_j G_{c,t}$, and a finite activity part, $X_{j,t} +p_jX_{c,t}$. Furthermore, all the margins at time one follow a generalized gamma law and we denote this writing  $\hat{G}_{j,1}\sim \hat{\Gamma}\left(1-k_j, q_j, k_j, p_j\right)$.
The joint characteristic function of $\hat{G} = \{\hat{G}_t, t\geq 0\}$ is given by\\
\begin{equation}\label{eq:CfMGG} 
\begin{split}
\Psi_{\hat{G}_{t}}\left(u\right)&\;= \prod_{j=1}^n \Psi_{G_{j,t}}\left(u_j\right)\Psi_{X_{j,t}}\left(u_j\right)\Psi_{G_{c,t}}\left(\sum_{j=1}^n u_j q_j\right)\Psi_{X_{c,t}}\left(\sum_{j=1}^n u_j p_j\right)\\ &\;\quad = \prod_{j=1}^n \exp \left(\frac{itu_j\left(k_j-c_2p_j\right)}{1-iu_jp_j}\right) \left( 1-iu_j q_j\right)^{t\left(\frac{k_j-1}{q_j}+c_1\right)} \\ &\;\quad \exp \left(\frac{i t c_2 \sum_{j=1}^n u_jp_j }{1- i\sum_{j=1}^n  u_jp_j}\right) \left(1- i \sum_{j=1}^n u_j q_j\right)^{-tc_1}.
\end{split}
\end{equation}
Setting $u_i=0$, $\forall i\neq j$, into (\ref{eq:CfMGG}) we get the characteristic function of $\hat{G}_j = \{\hat{G}_{j,t}, t\geq 0\}$
\begin{equation*}
\begin{split}
\Psi_{\hat{G}_{j,t}}\left(u_j\right)&\;= \exp \left(\frac{iu_jtk_j}{1-iu_jp_j}\right) \left( 1-iu_j q_j\right)^{t\frac{k_j-1}{q_j}}.
\end{split}
\end{equation*}

\cite{marfe2012generalized} defined the multivariate generalized variance gamma (MGVG) process as the L\'evy process $Y = \{Y_{t}, t\geq 0\}$ on $\R^n$ obtained as the difference of two independent MGG processes. This construction allows to model separately the dependence in positive and negative jumps.

Hovewer, \cite{marfe2012generalized} provided an alternative way to build a MGVG process through subordination using the MGG process as a subordinator. We analyse in details only this second approach. Let $Y = \{Y_t, t\geq 0\}$ be a multivariate process such that the following equalities in law hold  
\begin{equation*}
\begin{split}
Y_t &\;=\mu t + Y_{t}^{1,I} + Y_{t}^{2,I} + Y_{t}^{1,\rho} + Y_{t}^{2,\rho}\\ 
&\;= \mu t + B_{G_{t}}^{1,I}+ B_{X_{t}}^{2,I} + B_{G_{c,t}}^{1,\rho}+ B_{X_{c,t}}^{2,\rho},
\end{split}
\end{equation*}
where
\begin{itemize}
\item $Y^{1,I} = \{Y_t^{1,I}, t\geq 0\}$ is constructed by subordinating an $n$-dimensional arithmetic Brownian motion $B^{1,I}=\{B_t^{1,I},t\geq 0\}$ with an $n$-dimensional subordinator $G = \{G_{t}, t\geq 0\}$ with independent components $G_j = \{G_{j,t}, t\geq 0\}$, i.e $B_G^{1,I} = \{D_{G_{t}} \theta+D_{\sigma} W_{G_{t}}^{1,I}, t\geq 0\}$;
\item $Y^{2,I} = \{Y_t^{2,I}, t\geq 0\}$ is constructed by subordinating an $n$-dimensional arithmetic Brownian motion $B^{2,I}=\{B_t^{2,I},t\geq 0\}$  with an $n$-dimensional subordinator $X = \{X_{t}, t\geq 0\}$ with independent components $X_j = \{X_{j,t}, t\geq 0\}$, i.e $B_X^{2,I} = \{D_{X_{t}} \theta+D_{\sigma} W_{X_{t}}^{2,I}, t\geq 0\}$;
\item $Y^{1,\rho} = \{Y_t^{1,\rho}, t\geq 0\}$ is constructed by subordinating an $n$-dimensional Brownian motion $B^{1, \rho}=\{B^{1,\rho}_t, t\geq 0\}$ with the common $one$-dimensional subordinator $G_c = \{G_{c,t}, t\geq 0\}$, i.e $B_{G_c}^{1,\rho} = \{D_{G_{c,t}} \theta^{q}+D_{\sigma^{q}} W_{G_{c,t}}^{1,\rho}, t\geq 0\}$;
\item  $Y^{2,\rho} = \{Y_t^{2,\rho}, t\geq 0\}$ is constructed by subordinating an $n$-dimensional Brownian motion $B^{2, \rho}=\{B^{2,\rho}_t, t\geq 0\}$ with the common $one$-dimensional subordinator $X_c = \{X_{c,t}, t\geq 0\}$, i.e $B_{X_c}^{2,\rho} = \{D_{X_{c,t}} \theta^{p}+D_{\sigma^{p}} W_{X_{c,t}}^{2,\rho}, t\geq 0\}$;
\item $W^{l,I} = \{W_t^{l,I}, t\geq 0\}$ and $W^{l,\rho} = \{W_t^{l,\rho}, t\geq 0\}$ are independent $n$-dimensional Wiener processes, with $corr \left[  {W_{j,t}^{l,I},W_{k,t}^{l,I}}\right] =0$ and $corr \left[  {W_{j,t}^{l,\rho},W_{k,t}^{l,\rho}}\right] =\rho_{jk} $ for $j\neq k$ and $l=1, 2$;
\item $G$, $X$, $G_{c}$ and $X_{c}$ are independent subordinators, independent of $W^{l,I}$ and $W^{l,\rho}$ for $l=1, 2$;
\item $D_{G_{t}}$, $D_{X_{t}}$, $D_{G_{c,t}}$, $D_{X_{c, t}}$, $D_{\sigma}$, $D_{\sigma^ {q}}$ and  $D_{\sigma^ {p}}$ are diagonal matrices with diagonal elements $G_{j,t}$, $X_{j,t}$, $G_{c,t}$, $X_{c,t}$, $\sigma_j$, $\sigma_j\sqrt{q_j}$ and $\sigma_j\sqrt{p_j}$, respectively, with $\sigma_j$, $q_j$ and $p_j$ $\in\R_{+}$ for all $j$.
\item $\mu, \theta$, $\theta^{q}$ and $\theta^{p}$ are vectors in $\R^n$ with $\theta^{q} = \theta\times q$, $\theta^{p} = \theta\times p$  where $q$ and $p\in\R_{0^+}^n$.
\end{itemize} 

From independence and following (\ref{eq:ChfFactorBased}) we get the characteristic function of the MGVG process 
\begin{equation}\label{eq:CfMGVG} 
\begin{split}
\Psi_{Y_{t}}\left(u\right)&\;= \prod_{j=1}^n \left( 1-q_j\left( iu_j\theta_j - \frac{1}{2}u_j^2\sigma_j^2\right)\right)^{t\left(\frac{k_j-1}{q_j}+c_1\right)}\exp \left(\frac{t\left(k_j-c_2p_j\right)\left( iu_j\theta_j - \frac{1}{2}u_j^2\sigma_j^2\right)}{1-p_j\left( iu_j\theta_j - \frac{1}{2}u_j^2\sigma_j^2\right)}\right)  \\ &\;\quad \left(1-\left(i u'\theta^{q}-\frac{1}{2} u'\Sigma^{q} u\right)\right)^{-tc_1}\exp \left( t\left(iu'\mu+\frac{c_2 \left(i u'\theta^{p}-\frac{1}{2} u'\Sigma^{p} u\right)}{1-\left(i u'\theta^{p}-\frac{1}{2} u'\Sigma^{p} u\right)}\right)\right),
\end{split}
\end{equation}
where $\Sigma^{q}=Var\left(B_1^{1,\rho}\right)$ and $\Sigma^{p}=Var\left(B_1^{2,\rho}\right)$ are positive definite matrices with elements $\Sigma_{jk}^{q}=\sigma_j \sigma_k \sqrt{q_j} \sqrt{q_k} \rho_{jk}$ and  $\Sigma_{jk}^{p}=\sigma_j \sigma_k \sqrt{p_j} \sqrt{p_k} \rho_{jk}$. 
Setting $u_i=0$, $\forall i\neq j$, into (\ref{eq:CfMGVG}) we get the characteristic function of the  $j$-th marginal GVG process
\begin{equation*}
\begin{split}
\Psi_{Y_{j,t}}\left(u\right)&\;= \left( 1-q_j\left( iu_j\theta_j - \frac{1}{2}u_j^2\sigma_j^2\right)\right)^{t\frac{k_j-1}{q_j}} \exp\left(t\left(iu_j\mu_j+\frac{k_j \left( iu_j\theta_j - \frac{1}{2}u_j^2\sigma_j^2\right)}{1-p_j\left( iu_j\theta_j - \frac{1}{2}u_j^2\sigma_j^2\right)}\right)\right).
\end{split}
\end{equation*}

\cite{marfe2012generalized} suggested to estimate the MGVG process in two steps. First, estimate the margins with maximum likelihood estimation, recovering the density function from the characteristic function by using the FFT algorithm. Then, given the estimates of margins, estimate the common parameters to calibrate the empirical correlations or the empirical co-skewnesses or both at the same time.

\section{Linear combination of L\'evy processes}\label{linear}

In this section we discuss how to construct multivariate L\'evy models using affine linear transformations of random vectors with independent L\'evy components as proposed in \cite{kawai2009multivariate}, \cite{kaishev2013levy} and further studied by \cite{ballotta2016multivariate} and \cite{loregian2018estimation}. These approaches are based on the independent component analysis (ICA) and the principal component analysis (PCA).

\subsection{ICA based multivariate linear models}\label{sec:IcaLinear}

The idea behind this approach is to find a matrix $A \in \mathbb{R}^{n\times n}$ and a random vector $X=(X_1,\ldots,X_n)'$ with infinitely divisible, independent, and standardized components such that the law of the vector $AX$ approximates the law of the standardized log-returns while the correlation matrix of $AX$ approximates a given correlation matrix.

We define a new random vector $Z$ with $n$ entries as follows
\begin{equation}\label{eq:LinearAffine}
Z=AX+b,
\end{equation}
where $b\in \mathbb{R}^n$. Requiring that $X$ is a square integrable random vector and assuming, without loss of generality, $E\left[  XX'\right] = I_n$ with $E\left[ X\right] =0$, we have that the following equality holds
\begin{equation*}
var\left[ Z\right]  = AA '.
\end{equation*}
If $X$ is an infinitely divisible random vector, we have that $Z$ inherits this property from which it is possible to determine its L\'evy measure and the associated characteristic function. The corresponding L\'evy process $Z=\left\lbrace Z_t, t\geq 0\right\rbrace $ is defined as follows:
\begin{equation}\label{eq:LevProcAffineLin}
Z_t = AX_t + bt.
\end{equation}
Assuming that each component of the vector $X$ is not normally distributed it is possible to separate the estimation of matrix $A$ from the estimation of parameters of each component in $X$, through the ICA proposed in \cite{Comon94}. In the ICA approach, the dependence structure of the components in the vector $Z$ is described through the matrix $A$, called mixing matrix, that can be easily computed using the {\it FastICA} algorithm developed in \cite{Hyv2000}. Through this approach, \cite{madan2004asset} developed a multivariate VG model for asset returns and introduced a portfolio selection procedure based on the maximization of the expected CARA utility function. This approach has been further investigated in \cite{HITAJ2015146} and \cite{Mercuri2018}, where the components of the vector $X$ are assumed to be independent and mixed tempered stable distributed. 

In the following, we discuss two alternative approaches for constructing multivariate L\'evy models through a scheme described in equation \eqref{eq:LinearAffine}. The first method proposed in \cite{kawai2009multivariate} and based on the CTS distribution, and the second one proposed in \cite{kaishev2013levy} in which a multivariate L\'evy process is built as a linear combination of independent gamma processes. 

In both cases analyzed in Sections \ref{sec:LinearCTS} and \ref{sec:LinearLG}, we first standardize the margins and then we apply the {\it FastICA} algorithm to find the independent components of the vector $X$. For each margin $j$ we have
\begin{equation}\label{eq:Linear}
\frac{Y_j - \mu_j}{\sigma_j} = Z_j = A_jX + b_j,
\end{equation}
where $\mu_j$ and $\sigma_j$ are the empirical mean and empirical standard deviation of $Y_j$, $b_j$ is equal to zero and $A_j$ is the $j$-th row of the matrix $A$. We assume that theoretical means and standard deviations are estimated without errors. By construction means, standard deviations and correlation matrix of the model in equation (\ref{eq:Linear}) correspond with the empirical ones. The equation (\ref{eq:Linear}) allows one to obtain the characteristic function of $Y_j$ given the estimates of the standardized independent component $X$. The characteristic function of the linear combination in equation (\ref{eq:LevProcAffineLin}) can be written as follows
\begin{equation*}\label{eq:chfLevProcAffineLin}
\Psi_{Z_t}(u) = \exp(iu'bt)\Psi_{X_t}(A'u).
\end{equation*}

Additionally, given the moments of the independent components, it is possible to compute the moments of the original margin $Y_j$, for $j$ from $1$ to $n$. From the homogeneity property and the additivity property of independent random variables, it follows that the cumulant of order $k$ of a linear combination of independent random variables is a linear combination of their cumulants of the same order with coefficients raised to the power $k$. 

Thus, these models can be estimated by considering the moment matching approach as described in Section \ref{sec:BruteForce} or by applying a maximum likelihood estimation on each standardized univariate independent component, and by computing, then, the density of each margin $Y_j$ as linear combination of the independent standardized components.

\subsubsection{Multivariate linear classical tempered stable model}\label{sec:LinearCTS}

As proposed in \cite{kawai2009multivariate}, assume that $X=\left\{X_t, t\geq 0\right\} $ is a L\'evy process in $\mathbb{R}^n$ without a Gaussian component and 
\begin{equation}\label{varianceRestr}
var\left[ X_{1,t}\right] =\ldots=var\left[ X_{n,t}\right]=t\xi^2
\end{equation}
holds, with $\xi>0$. The $j$-th component of the stochastic process $Y=\left\{Y_t, t\geq0\right\}$ is defined by the  equality
\begin{equation*}
Y_{j,t}=\sum_{l=1}^n c_{j,l}X_{l,t}.
\end{equation*}
For a fixed correlation matrix $\mho$, the transformation matrix $K$ such that $KK'=\mho$ can be obtained for example through a singular value decomposition. The restriction \eqref{varianceRestr} implies additional constraints on the marginal parameters of $X_{j,t}$ during the calibration procedure. In order to be in the Kawai's framework and to avoid this additional constraint in the calibration algorithm, we apply the {\it FastICA} algorithm to the standardized multivariate returns as in equation (\ref{eq:Linear}). Then on each independent component $j$ we estimate a univariate standardized tempered stable model having the following characteristic function with parameters ($\alpha_j$, $\lambda_{j+}$, $\lambda_{j-}$)
\begin{equation*}
\begin{split}
\phi_{X_j}(u) = E[\exp(iuX_j)] = &\; \exp\Big(-iu\big(C\Gamma(1 - \alpha_j)(\lambda_{j+}^{\alpha_j - 1} - \lambda_{j-}^{\alpha_j - 1})\big)\\           
&\; + C\Gamma(-\alpha_j)((\lambda_{j+} - iu)^{\alpha_j} - \lambda_{j+}^{\alpha_j} + (\lambda_{j-} + iu)^{\alpha_j}  - \lambda_{j-}^{\alpha_j})\Big)
\end{split}
\end{equation*}
where 
$$
C = (\Gamma(2 - \alpha_j)(\lambda_{j+}^{\alpha_j - 2} + \lambda_{j-}^{\alpha_j - 2}))^{-1},
$$
and the cumulants are $c_1(X_j)=0$, $c_2(X_j)$ = 1, 
$$
c_3(X_j) = C\Gamma(3 - \alpha_j)(\lambda_{j+}^{\alpha_j - 3} - \lambda_{j-}^{\alpha_j - 3}),
$$
and
$$
c_4(X_j) = C\Gamma(4 - \alpha_j)(\lambda_{j+}^{\alpha_j - 4} - \lambda_{-}^{\alpha_j - 4}).
$$
We refer to this multivariate L\'evy process built as linear combination of independent CTS processes as multivariate linear classical tempered stable (MLCTS) model. While \cite{kawai2009multivariate} considered a process under the so-called mean-correcting martingale measure, we estimate the model under the historical measure where the mean of each margins corresponds to the empirical one.

\subsubsection{Multivariate linear gamma model}\label{sec:LinearLG}

\cite{kaishev2013levy} proposed a new class of processes defined as linear combination of independent gamma processes, called LG processes. In this paper we consider and estimate a special case of this processes, where the univariate standardized independent components are define as sum of independent gamma processes, that is:
\begin{equation*}
X_j = \sum_{k=1}^d d_k G_{k,t},
\end{equation*}
where $G_{k,t}\sim\Gamma\left(a_k t, \lambda\right)$. We refer to it as multivariate linear gamma (MLG) model. We assume $d=2$  and on each independent component $j$ we estimate a univariate linear gamma model having the following characteristic function with parameters ($\lambda$, $a_{j+}$, $a_{j-}$)

\begin{equation*}
\begin{split}
\phi_{X_j}(u) = E[\exp(iuX_j)] = &\; \exp\Big(a_{j+}\log(\lambda) - a_{j+}\log(\lambda - id_{j+}u)\\ 
& + a_{j-}\log(\lambda) - a_{j-}\log(\lambda + id_{j+}u)\Big),
\end{split}
\end{equation*}
where  
\begin{equation*}
\begin{split}
  d_{j+} =& \lambda\sqrt{\frac{a_{j-}}{a_{j+}(a_{j+} + a_{j-})}}, \\
  d_{j-} =& \lambda\sqrt{\frac{a_{j+}}{a_{j-}(a_{j+} + a_{j-})}},
\end{split}
\end{equation*}
and the cumulants are $c_1(X_j)=0$, $c_2(X_j)$ = 1, 
$$
c_3(X_j) = 2{\lambda}^{-3}({a_{j+}d_{j+}^3} - {a_{j-}d_{j-}^3})
$$
and
$$
c_4(X_j) = 6{\lambda}^{-4}({a_{j+}d_{j+}^4} + {a_{j-}d_{j-}^4}).
$$

\subsection{PCA based multivariate linear models}\label{sec:LinearCTS_Ballotta}

A further approach to build multivariate models based on linear combination of independent L\'evy processes has been recently proposed by \cite{ballotta2016multivariate} and \cite{loregian2018estimation}. This approach can be viewed as a further extension of the methods described in  Section \ref{linear}, even if \cite{loregian2018estimation} proposed an estimation approach based on the principal component analysis (PCA). 

For the case of an $n$-dimensional model, the authors suggested a 2-step estimation procedure in which a common factor $\Upsilon$ has to be estimated first and then $n$ univariate estimations should be conducted, one per each idiosyncratic component. The model is defined as linear combination of two independent L\'evy processes, the first representing a common risk component, the second representing the idiosyncratic risks. The first component is the first principal component defined through the PCA. Even if it is possible to extend the model to the first $k$ principal components to capture $k$ different common risk factors, we consider only the first one in the empirical application. 

Let $X_j =\{X_{j,t}, t>=0\}$ and $\Upsilon=\{\Upsilon_{t}, t>=0\}$ be two independent L\'evy processes belonging to the same parametric family (e.g. CTS L\'evy processes with possible different parameters), then $Y_j =\{Y_{j,t}, t>=0\}$ can be defined as follows
\begin{equation*}
Y_j = X_j + f_j\Upsilon,
\end{equation*}
where $f_j$ is the $j$-th component of the vector $f\in\R^n$. Given the characteristic functions of $X_j$ and $\Upsilon$, it is simple to obtain the characteristic function of $Y_j$ as well as its cumulants. According to equation (3) in \cite{loregian2018estimation}, it can be shown that the correlation is given by the following formula
\begin{equation*}
corr(Y_j,Y_k) = \frac{f_jf_k var[\Upsilon]}{\sqrt{var[Y_j]}\sqrt{var[Y_k]}}.
\end{equation*}
The estimation of this model can be conducted by maximum likelihood estimation (MLE) through the FFT: it is fast to implement, and its complexity does not increase with the number of components of the multivariate model. First, the parameters of $\Upsilon$ are estimated, then $n$ independent MLE are performed to estimate the parameters of $X_j$ and the vector $f$. By construction, this second step can be parallelized in a straightforward way. While \cite{loregian2018estimation} conducted the empirical analysis by considering NIG and Merton jump-diffusion processes, we assume that the risk components are CTS distributed. We refer to this multivariate L\'evy process built as linear combination of independent CTS processes as multivariate linear classical tempered stable (MLCTS) model.

\section{Estimation methods}\label{estimMethod}

From a theoretical standpoint, a good estimator should satisfy the following properties: (1) the expected value of the estimator should be equal to the true value of the parameter ({\it unbiasedness}); (2) as the dimension of the sample increases the estimator should converge in probability to the true value of the parameter ({\it consistency}); (3) among the unbiased estimators the selected one should be that with the smallest variance ({\it efficiency}). The knowledge of the sample distribution of an estimator allows to perform hypothesis testing on model parameters. In this section we discuss different estimation methods used in the literature. For each method, we explain the underlying theoretical requirements that ensure proper statistical properties of estimators.

\subsection{Moments matching (or brute force)}\label{sec:BruteForce}

The knowledge of the characteristic function of a multivariate parametric model allows to derive the theoretical moments of the margins and of the joint distribution. Since theoretical moments are expressed as a function of the unknown parameters, it is possible to estimate model parameters by minimizing the distance between empirical and theoretical moments. This simple approach can be applied to estimate all multivariate models  discussed in this contribution. We refer to this sort of moments matching estimation method as the {\it brute force} approach. More in details we minimize the Euclidean norm of the difference between the first four empirical and theoretical marginal moments and the Frobenius norm of the difference between empirical and theoretical correlation matrices, that is
\begin{equation}\label{eq:BruteForce}
 \min_{\Theta} \left(\sum_{i=1}^4 w_i\|m^*_i - m_i(\Theta)\| + w_{\rho}\|\rho^*_i - \rho(\Theta)\|_F\right),
\end{equation}
where $m^*_i$ and $m_i(\Theta)$ are the empirical and the theoretical marginal moments of order $i$, $\rho^*$ and $\rho(\Theta)$ are the empirical and theoretical correlation matrices, $w_i$ and $w_{\rho}$ are weights.

To take into account the characteristics of each model, including the number of parameters, and to avoid numerical errors in the optimization algorithm, we use different weights $w_i$ and $w_{\rho}$. Since this type of approach strictly depends on the starting point in the optimization algorithm, we randomly draw 100 different starting points and select as result the point of minimum distance among the 100 solutions. A careful selection of both the upper and lower parameters bounds is needed to have a satisfactory performance of the optimization algorithm. Since the theoretical moments have a closed form formula, the algorithm is fast for all models. A similar approach will be used in the two-step approach described in Section \ref{sec:TwoStep} to minimize the distance between empirical and theoretical correlation matrices to find the common parameters governing the dependence structure.

It should be noted that for some models to ensure  a proper correlation matrix we apply the hypersphere decomposition as described in \cite{rebonato2011most}, that is the correlation matrix of dimension $n$ is decomposed as the product of a lower triangular matrix $B$ and its transpose $B'$. This lower triangular matrix is function of $n(n-1)/2$ angles and $BB'$ is by construction a correlation matrix. 

\subsection{Maximum likelihood estimation}\label{mle}

Let us consider a multivariate random vector, that is a random variable $Y$ that assumes values on $\mathbb{R}^n$ with an assigned probability law. Given a set of $T$ observations $\{Y^k  = Y_{t_k} - Y_{t_{k-1}}\}_{k=1,\ldots,T}$, the log-likelihood function can be written as
\begin{equation}\label{eq:LL}
\begin{split}
LL(\Theta;Y^1,\ldots,Y^T) &\;= \sum_{k=1}^T \log f_Y(Y^k;\Theta),
\end{split}
\end{equation} 
where $\Theta$ is the set of parameters. The idea behind the MLE is to choose the vector $\Theta$ that maximizes the likelihood, or equivalently, the logarithm of the likelihood of the observed sample, that is
\begin{equation}\label{eq:MLE}
\max_{\Theta} LL(\Theta;Y^1,\ldots,Y^T).
\end{equation}
Under mild conditions, the method ensures the consistency property while the efficiency is attained only asymptotically. Moreover, estimators converge in law to the Gaussian distribution with rate $\frac{1}{\sqrt{T}}$. The procedure can be used if it exists a closed form formula of the joint density function. However, as soon as the dimension increases, the optimization problem in equation (\ref{eq:MLE}) becomes infeasible. 

We will consider the MLE algorithm only in the univariate case to estimate the parameter of the margins (e.g. in the two-step procedure described in Section \ref{sec:TwoStep} or for estimating the linear models described in Section \ref{linear}). In the GH case there is a closed form formula for the density function, therefore the likelihood function is simple to compute. In all other cases we will compute the density function by means of the FFT as discussed in details in \cite{bstff2019handbook}.

\subsection{Expectation maximization MLE method}\label{sec:EM}

In the subclass of multivariate infinitely divisible distributions that can be written as a mixture, the maximum likelihood approach can be performed using the expectation maximization (EM) algorithm proposed by \cite{Dempster1977}. In the class of normal mean-variance mixtures it is necessary to be able to evaluate the posterior distribution of the mixing random variable. For univariate distributions explicit expressions for estimators of parameters have been given in \cite{Dimitris2002} for the univariate NIG and in \cite{Loregian2012} for the univariate VG. \cite{liu1994ecme}, \cite{hu2005calibration} and \cite{embrechts2005quantitative} study the EM-based maximum likelihood algorithm for estimating the parameters of the MGH distribution. \cite{btf2015riding} proposed a simple expectation-maximization maximum likelihood estimation procedure for the MNTS model where the density function of the mixing random variable is computed by means of a FFT procedure.

The density function of a normal mean-variance mixtures distribution can be written as
\begin{equation}\label{eq:densityNMVM}
f_Y(y;\Theta) = \int_0^\infty f_{Y|S}(y|s; \mu,\theta,\Sigma)h(s;\Theta_h)ds,
\end{equation}
where $Y|S\sim N(\mu + \theta S,S\Sigma)$ (see \cite{hu2005calibration}), $h$ is the density function of the mixing random variable with parameter set $\Theta_h$ (e.g. the  set of parameters of the GIG distribution in the MGH case), and $\Theta$ is the set of all model parameters. In the MNTS case the density function $h$ is computed by means of a FFT procedure, that is the characteristic function is inverted to calculate the density function $h$ and the density $f_Y$ in equation (\ref{eq:densityNMVM}) has to be found by numerical integration.

Given a set of $T$ observations $\{Y^k  = Y_{t_k} - Y_{t_{k-1}}\}_{k=1,\ldots,T}$, the log-likelihood can be written as
\begin{equation}\label{eq:LL1}
\begin{split}
LL(\Theta;Y^1,\ldots,Y^T) &\;= \sum_{k=1}^T \log f_Y(Y^k;\Theta).
\end{split}
\end{equation} 
We consider the following likelihood function instead of the likelihood in equation (\ref{eq:LL1})  
\begin{equation}\label{eq:LL2}
\begin{split}
LL(\Theta;Y^1,\ldots,Y^T, S^1,\ldots,S^T) &\;= \sum_{k=1}^T \log f_{Y,S}(Y^k,S^k;\Theta)\\
&\;= \sum_{k=1}^T \log f_{Y|S}(Y^k|S^k;\mu,\theta,\Sigma) + \sum_{k=1}^T \log h_{S}(S^k;\Theta_h)\\
&\;= L_1(\mu,\theta,\Sigma; Y|S) + L_2(\Theta_h; S),
\end{split}
\end{equation} 
where $\{S^k = S_{t_k} - S_{t_{k-1}}\}_{k=1,\ldots,T}$ the latent mixing variables. In order to find a MLE based on (\ref{eq:LL2}), we consider the following iterative algorithm. 

\begin{enumerate}
 \item Set $i=1$ and select a starting value for $\Theta^{(1)}$, that is $\mu^{(1)}\in\R^n$ is the sample mean, $\theta^{(1)}\in\R^n$ is the zero vector, $V\in\R^n\times\R^n$ is the sample covariance matrix.
 \item By considering that 
\begin{equation}\label{eq:ratio}
f_{S|Y^k}(s; Y^k, \Theta) = \frac{f_{Y|S}(Y^k|s; \mu,\theta,\Sigma)h(s;\Theta_h)}{f_Y(Y^k;\Theta)},
\end{equation}
compute the following weights
\begin{equation}\label{eq:weights}
\begin{split}
\delta_k^{(\cdot)} &\;= E({S^k}^{-1}|Y^k, \Theta^{(\cdot)}),\\ 
\eta_k^{(\cdot)} &\;= E({S^k}|Y^k, \Theta^{(\cdot)}),\\ 
\rho^{(i)}_k &\;= (Y^k - \mu^{(i)})'\left(\Sigma^{(i)}\right)^{-1}(Y^k - \mu^{(i)}),
\end{split}
\end{equation}
The expectations in equation (\ref{eq:weights}) are evaluated by numerical integration. In equation (\ref{eq:ratio}), $f_{Y|S}$ can be written in closed form since $Y|S^k\sim N(\mu + \theta S^k,S^k\Sigma)$ (see \cite{hu2005calibration}). While in the MNTS case the density $h$ is computed by means of a FFT procedure, and the denominator is evaluated by numerical integration, in the MGH case both functions have a closed form formula.
\item Evaluate the average values
$$
\bar{\delta}^{(i)}=\sum_{k=1}^T\delta_k^{(i)}, \qquad \bar{\eta}^{(i)}=\sum_{k=1}^T\eta_k^{(i)}.
$$
%
\item Get the estimates
\begin{equation*}
\begin{split}
\theta^{(i+1)} &\;= \frac{N^{-1}\sum_{k=1}^T\delta_k^{(i)}(\bar{Y} - Y^k)}{\bar{\delta}^{(i)}\bar{\eta}^{(i)}-1},\\
\end{split}
\end{equation*} 
\begin{equation*}
\begin{split}
\mu^{(i+1)} &\;= \frac{N^{-1}\sum_{k=1}^T\delta_k^{(i)}Y^k - \theta^{(i+1)}}{\bar{\delta}^{(i)}},\\
\Psi &\;= \frac{1}{N} \sum_{k=1}^T \delta_k^{(i)} (Y^k - \mu^{(i+1)})(Y^k - \mu^{(i+1)})' - \bar{\eta}^{(i)} \theta^{(i+1)} {\theta^{(i+1)}}',\\
\Sigma^{(i+1)}&\;=\frac{|V|^{1/n}\Psi}{|\Psi|^{1/n}}.
\end{split}
\end{equation*} 
\item Set $\Theta^{(i')} = \{\Theta_h^{(i)},  \theta^{(i+1)}, \mu^{(i + 1)}, \Sigma^{(i+1)}\}$ and calculate the new weight $\bar{\eta}^{(i')}$ as done in Steps 2 and 3.
\item To complete the calculation of $\Theta^{(i+1)}$, find $\Theta_h$ that maximize the likelihood function in equation (\ref{eq:LL1}), that is
\begin{equation*}
LL(\Theta^{(i+1)};Y^1,\ldots,Y^T) = \sum_{k=1}^T \log f_Y(Y^k;\Theta^{(i+1)}),
\end{equation*} 
where $\Theta^{(i+1)} = \{a, \lambda, C, \theta^{(i+1)}, \mu^{(i + 1)}, \Sigma^{(i+1)}\}$. 
\item If $i < 1,000$ and $LL(i) - LL(i-1) > 1e-5$, increment iteration count $i$ and go to step 2, otherwise, stop the algorithm.
\end{enumerate}
As discussed in \cite{embrechts2005quantitative}, the algorithm converges to the MLE because it produces improved parameter estimates at each step, in the sense that the value of the original likelihood is continually increased. In the applications contained in this paper, the algorithm converges long before reaching 1,000 iterations.

\subsection{Two-step procedure}\label{sec:TwoStep}

In literature several authors have considered a two-step procedure for the estimation of non-Gaussian multivariate models. The idea is to split parameters into two groups: the parameters of the first group can be estimated on the margins, usually by MLE, while the parameters of the second group, used to capture the dependence structure, can be estimated by minimizing some distance between the theoretical and empirical higher co-moments. While marginal parameters are estimated using the MLE method discussed in Section \ref{mle}, the Frobenius norm is usually applied to minimize the distance between theoretical and empirical correlation matrices (see Section \ref{sec:BruteForce}). A two-step procedure was used in \cite{marfe2012generalized}, \cite{hitaj2013portfolio},  \cite{luciano2016dependence} and \cite{boen2019building}. 

Here we discuss the conditions required for applying this method and statistical properties of estimators. If it is not possible to identify the set of parameters that completely characterize the margins this method should not be applied. For example this is the case of NMV as the parameters governing the dependence structure affect all margins. A situation when the two-step procedure can be applied refers to the case where the dependence structure is introduced through a multivariate mixing random variable as for example in \cite{luciano2010multivariate}. Notice that it is necessary to impose constraints on the parameters of the subordinator leading to marginal distributions which become functions only of parameters of the first group (see \cite{Guillaume2013} for a discussion about these constraints in case of multivariate $\alpha$VG and $\alpha$NIG). 
 
\subsection{Estimation methods based on the characteristic function}\label{gmm}

Most of the multivariate models reviewed in this paper do not have a closed form formula for the density function. This makes the likelihood-based estimation inconvenient. However, these models can be represented through some transformations of the probability function such as the characteristic function or the Laplace transform.
We discuss briefly the generalized  method of moments (GMM) proposed in \cite{Hansen1982} that can be used for the estimation of all the multivariate models described in this paper. The aim of this procedure is to estimate the  vector of parameters $\Theta \in \mathbb{R}^p$ from a model based on the following vector of $q$ unconditional moment conditions:
\begin{equation}
E\left[g\left(\Theta,Y\right)\right]=0
\label{gmmexp}
\end{equation}
where $g\left(\cdot\right):\Xi \times\mathbb{R}^{n} \rightarrow \mathbb{R}^{q}$, $\Xi \subset \mathbb{R}^p$ is a compact space. 
For a given sample $Y^1,Y^2,\ldots,Y^T$, we replace the expectation in \eqref{gmmexp} with the sample mean and obtain:
\begin{equation*}
\bar{g}_T\left(\Theta\right)=\frac{1}{T}\sum_{k=1}^T g\left(\Theta,Y^k\right).
\end{equation*}
The GMM estimator depends on the choice of a positive definite weighting matrix $F_T\in \mathbb{R}^{q\times q}$  
and is the solution of the following minimization problem
\begin{equation}\label{estTheta}
\hat{\Theta}\left(F_T\right)=\min_{\Theta \in \Xi} \ \bar{g}_T\left(\Theta\right)'F_T \bar{g}_T\left(\Theta\right).
\end{equation}
To find an estimator, we need at least as many moment conditions as the number of parameters. In particular we have the classical method of moments (MM) for $q = p$ and the GMM for $q>p$. Under mild conditions, for any positive definite weighting matrix $F_T$, the GMM produces consistent estimators. Moreover the asymptotic distribution of $\hat{\Theta}$ is
\[
\sqrt{T}\left(\hat{\Theta}-\Theta\right)\sim N\left(0,H\right)
\] 
where $H$ is the asymptotic variance-covariance matrix defined as
\[
H = \left(J' F J\right)^{-1} J' F R F J \left(J' F J\right)^{-1}
\]  
with $J \in \mathbb{R}^{q\times p}$
\[
J = E\left[\frac{\partial g\left(\Theta,Y\right)}{\partial \Theta'}\right]
\]
and $R \in \mathbb{R}^{q\times q}$
\[
R=E\left[g\left(\Theta,Y\right)g\left(\Theta,Y\right)'\right],
\]
and $F_T\overset{P}{\to} F$ as $T\rightarrow\infty$. An appropriate choice of matrix $F$ improves the efficiency within the class of GMM type estimators. The most efficient estimator is obtained if 
\begin{equation}\label{eq:WeightingMatrix}
F_T \overset{P}{\to}  R^{-1}
\end{equation}
and, in that case, the variance-covariance matrix $H$ becomes
\begin{equation}\label{V:eqr}
H = \left(J' R^{-1} J\right)^{-1}.
\end{equation}
Several algorithms have been proposed in literature in order to obtain an estimator with variance-covariance matrix that approaches matrix $H$ in \eqref{V:eqr}. 

In this paper we use the \textsf{R} package  \texttt{gmm} developed in \cite{chausse2010}, where the optimal matrix $F$ is estimated using the heteroskedastic auto-correlation consistent (HAC) approach proposed in \cite{Newey1987}. Then, as $g$ we select the distance between the empirical and theoretical characteristic function. For a given grid $u_j \in \mathbb{R}^{n}$, with $j$ from 1 to $q$, $g$ is defined as
\[
g\left(\Theta, Y, u_j\right) = e^{i\left\langle u_j,Y\right\rangle}-\Psi_{Y,\Theta}\left(u_j\right), 
\]
where $\left\langle \ , \ \right\rangle$ is the scalar product. The moment conditions require
\[
E\left[g\left(\Theta, Y, u_j\right)\right]=0,
\]
where the $j$-th element of the vector function $\bar{g}_T\left(\Theta\right)$ is
\begin{equation}\label{eq:gmm_gt}
\bar{g}_{T,j}\left(\Theta\right)=\frac{1}{T}\sum_{k=1}^T \left(e^{i\left\langle u_j, Y^k\right\rangle}-\Psi_{Y^k,\Theta}\left(u_j\right)\right).
\end{equation}
It is evident that the choice of the grid is crucial. Increasing the grid size ${u_j}$ implies numerical instability and $R^{-1}$ may not be defined. 

\section{Empirical analysis}\label{empan}

In this paper we have tried to highlight the differences between various multivariate distributions applied to finance. Also, different estimation methods have been discussed.  In order to select a {\it good} multivariate model it is necessary to understand the most important features we want to be able to replicate, the computational burden of the choice and the characteristics of the algorithm selected in the estimation. In this section we compare the different multivariate non-normal models with the multivariate normal one to which we refer to as {\it MNormal}. This model is completely characterized through the mean vector $\mu$ and covariance matrix $\Sigma$. 

In Table \ref{NuPaforMod} the number of parameters as a function of the number of margins is reported  for each multivariate model. The models with a linear dependence on the number of margins are the MMixedTS, the $\alpha$GH and the PCA based linear model with $k$ equal to $1$. For the other models the dependence is quadratic. 

In this section we report the estimation results of the models we discussed. We calibrate each model with three different algorithms (only the MMixedTS is estimated with two different approaches). We estimate the models under the so-called historical probability measure, that is by using observed market returns without relying on additional data, like for example option prices (see \cite{bianchi2017forward}).

The analysis is performed on Datastream daily dividend-adjusted closing prices from July 1, 2003 through June 29, 2018 for five stock indexes: the Deutsche Aktienindex 30 (ticker DAX),  the Cotation Assist\'ee en Continu 40 (ticker CAC),  the Financial Times Stock Exchange Milano Indice di Borsa (ticker FTSEMIB), \'Indice Burs\'atil Espa\~nol (ticker IBEX), Amsterdam Exchange Index (ticker AEX) representing five major European indexes. The time period in this study includes the high volatility period after the Lehman Brothers filing for Chapter 11 bankruptcy protection (September 15, 2008), the eurozone sovereign debt crisis, during which, in November 2011, the spread between the 10-year Italian BTP and the German Bund with the same maturity exceeded 500 basis points, and the recent Italian political turmoil at the end of May 2018.

\begin{table}
\centering
\begin{footnotesize}
\begin{tabular}{lc}
\toprule
Model & number of parameters \\ 
\midrule
MNormal &  $\frac{n^2+3n}{2}$\\
MGH  &  $\frac{n^2+5n}{2}+3$\\
MNTS &	$\frac{n^2+5n}{2}+3$\\
$\alpha$GH & $5n+2$ \\
$\alpha\rho$GH & $\frac{n^2+9n}{2}+2$ \\  
MMixedTS & $7n+1$ \\
MGVG & $\frac{n^2+11n}{2}+2$ \\
ICA MLG &  $n^2 + 5n$\\
ICA MLCTS & $n^2 + 5n$\\
PCA MLCTS & $5n+(n+5)k$\\
\bottomrule
\end{tabular}
\caption{\footnotesize Number of parameters as a function of the number of margins $n$. The number of principal components is equal to $k$.} 
\label{NuPaforMod}
\end{footnotesize}
\end{table}

We estimate the models using the methods discussed in Section \ref{estimMethod}. The code is implemented in R language. Three estimation error measures based on the distance between theoretical and empirical distribution function are considered.  The first error measure is the Kolmogorov-Smirnov distance ($KS$) 
\begin{equation}\label{eq:KS}
 KS=\sup_{x}\left|\hat{F}\left(x\right)-F\left(x\right)\right|,
\end{equation}
applied to all margins. If the margins do not have closed form formula for the density function, the evaluation of equation (\ref{eq:KS}) is conducted by means of the FTT as discussed in details in \cite{bstff2019handbook}. Given the number of observations, the KS distance provides a satisfactory result if it is less than 0.03 (i.e. the KS test has a $p$-value grater than 0.05). The second measure is represented by the distances between empirical and theoretical moments considered in Section \ref{sec:BruteForce}. The third error measure is given by the norm of the vector $\bar{g}_{T,j}$ defined in equation (\ref{eq:gmm_gt}), which gives the average distance between the empirical and the theoretical characteristic function given a grid $u_j \in \mathbb{R}^{n}$, with $j$ from 1 to 50. 

While Tables \ref{sumStatLog} reports the summary statistics for log-returns of the European indexes, in Table \ref{tab:EstimationError} we show the estimation errors in term of both margins, dependence structure and the whole multivariate distribution. As expected, even if the estimation method based on the moments (i.e. the brute force approach) is very simple to implement, it does not always provide a satisfactory estimation error. However, it can be a good starting point for the other estimation algorithms considered in this study. Due to the large number of parameters, in some cases it is not easy to understand which can be a good starting point for the optimization procedure. The brute force approach gives the possibility to explore a multivariate model without having to implement complex algorithms or without having to wait too long for the algorithm convergence. However, in some cases the estimation error is large if compared to more robust methods (e.g. the EM or the GMM algorithm). Even if the GMM algorithm can be applied to all models having a characteristic function in closed form, the computing time of this algorithm as well as possible numerical issues may be an obstacle for large scale practical applications. The GMM is a very general estimation approach and for this reason it may be more difficult to use in comparison with ad-hoc estimation approaches  implemented for specific cases. However, with a proper selection of the grid and of the parameter boundaries, the GMM provide satisfactories results.

\begin{table}
\centering
\begin{footnotesize}
\begin{tabular}{lcccccc}
  \hline
	&	min & max &  mean & std & skewness & ex.kurtosis \\
  \hline
  DAX &  -0.074 & 0.1080 & 0.0003 &  0.0131 &  -0.0510&  6.2406 \\ 
  CAC & -0.0947 & 0.1059 & 0.0001 & 0.0134 & -0.0458& 6.9923 \\
  FTSEMIB & -0.1333 & 0.1088 & -0.0000 & 0.0150 & -0.2547 &  6.1572 \\
  IBEX & -0.1319 &  0.1348 & 0.0001 & 0.0141 & -0.1459 &  8.3714 \\
  AEX & -0.0959 & 0.1003 & 0.0002 & 0.0126 & -0.1998& 8.8435\\
 \hline
\end{tabular}
\caption{\footnotesize Summary statistics of log-returns.}
\label{sumStatLog} 
\end{footnotesize}
\end{table}

As starting points of the moments-matching approach, we draw 100 random starting points in the parameters space. The $w_i$ and $w_{\rho}$ depend on the selected parametric model and the choice is done after exploring the dataset and the algorithm itself. We try to select the weights in a way that all moments have a similar importance in the optimization algorithm. For the MNTS, the MGH, the $\alpha$GH and the MLCTS model based on PCA we consider only mean, standard deviation and correlation in equation (\ref{eq:BruteForce}); for the $\rho\alpha$GH we consider also the skewness; for the MGVG and the MMixedTS we consider also both skewness and excess kurtosis; for the multivariate linear models based on ICA we consider all marginal moments up to order four. This is the reason why in Table \ref{tab:EstimationError} in both the MNTS and the MGH case the error in fitting the skewness and the excess kurtosis is large and in the other cases the error in fitting moments of order higher than two is not so big.

Both the EM and the two-step approaches are reasonably fast: the maximization of the likelihood function is conducted only on univariate models. The EM approach applied to the MGH and MNTS models works properly even for large scale practical applications (see \cite{bianchi2017forward} and \cite{bdr2020covar}). Our optimization algorithms in R do not rely on parallel computing techniques and the code implements the L-BFGS-B method. However, while it is not so simple to parallelize the optimization algorithm, it should be noted that the $n$ independent MLE steps of the models leveraging on univariate MLE can be run in parallel without great effort (e.g. it is possible to write an efficient R code with the packages {\it foreach} and {\it doParallel}). This is the case for the two-step approaches and the linear models (i.e. $\alpha$GH, $\rho\alpha$GH, MGVG, MLG and MLCTS) which are based on the {\it divide et impera} concept: the estimation procedure is simplified, the dimensionality problem is solved and the models provide a consistent and parallelizable parameters estimation. 

In the GMM estimation the multivariate grid of dimension $n$ is selected as follows. For the first dimension we consider a vector of $q$ equally spaced points in the interval between minimum and maximum observed returns. Then, after having fixed a seed, to obtain the vector representing the second dimension, we randomly permute the vector obtained for the first dimension. The same approach is considered for all other dimensions up to $n$. The value of $q$ depends on the model and it ranges from 15 to 50. As starting point we consider the estimates obtained through the moments-matching approach. First, we obtain a preliminary estimate by considering as weighting matrix the identity matrix, then we conduct a second estimation with the weighting matrix given in equation (\ref{eq:WeightingMatrix}). The selection of the starting point and of the grid largely affects the final result of the optimization procedure. The estimates obtained through the GMM approach are usually not far from the starting point. The GMM approach seems to work better for models with a simpler dependence structure and a smaller number of parameters (i.e. the MMixedTS model). A proper selection of $q$ is needed to avoid possible numerical issues of the R package {\it gmm}. Even if the GMM approach is reasonably fast, it is not always simple to obtain satisfactory results in terms of margins fitting and convergence properties. This may be caused by the large number of parameters involved in the optimization problem. In order to speed up the GMM algorithm and to avoid loops, the characteristic function should take as input a matrix $u$, instead of a vector $u$, and implement the code leveraging on matrix operations. This can be done for all parametric models analyzed in this work.

From an estimation error standpoint, some models have a very good performance in fitting the margins, but they show a bad correlation fitting (e.g. the MMixedTS and the MGVG). For the MMixedTS the large error in fitting the empirical correlation matrix seems to be due to the number of parameters, too small to explain the behavior of both margins and correlations. The MMixedTS is the best performer in fitting the margins. The MGH and the MNTS models have a satisfactory performance, even if the correlation fitting is not as good as for other competitors. At least for the data analyzed in this paper, the $\rho\alpha$GH seems to show the better mix between estimation errors and computational tractability, even if the two-step procedure is not elegant from a pure statistical perspective. The ICA based linear models are simple to estimate, mainly because the multivariate estimation problem is converted to a set of univariate problems. However, the performance is not always good enough and some numerical issues in the FFT inversion of the characteristic function may affect the evaluation of the estimation errors. These issues are caused by the fact that the model parameters are estimated on the independent components and the margins are obtained by multiplying these components by small numbers, that is by the elements of the matrix $A$ and of the vector $\sigma$ (see equation (\ref{eq:Linear})). The PCA based linear model with CTS components is more efficient from a computational standpoint and it is simpler to implement, at least if one considers the first principal component only. In our view, this last model is very promising, even if it has a less flexible dependence structure in comparison with the $\rho\alpha$GH. The estimation procedure has a computational complexity equivalent to the estimation of a non-normal univariate model.

\section{Conclusions}\label{sec:Conclusions}

In this paper we provide a guide for the use of multivariate non-Gaussian models with a view toward applications to finance. After a detailed analysis of the theoretical structure behind a sample of multivariate models proposed in the financial literature, we observed their performance in terms of fitting on a five-dimensional series of log-returns. The contribution of the paper is not only to present models with a unifying notation but also to give some inputs for the practical implementation of their estimation algorithms. For each model we provide the necessary formulas and methods needed to find a preliminary estimate that can be used as starting point of more complex and robust algorithms. Additionally, we propose different estimation methods which can be used in practical applications.

The parametric models reviewed is this paper have a different level of complexity from both a theoretical and practical standpoint. We show that it is not always true that a greater level of complexity provides a better estimation performance, at least for the data considered in this study. In most cases the multivariate estimation problem can be decomposed in different steps with computational complexity similar to a univariate estimation problem. When this decomposition is not possible, we show how to perfom a satisfactory parameters estimation.

As we expected, we are not able to identify a multivariate model that is more appropriate. Statistical properties of estimators and computational tractability are important features that should be taken into accout when selecting a model to be used in practice.

\begin{sidewaystable}
\begin{center}
\begin{scriptsize}
\begin{tabular}{@{}llcccccccccccc@{}}
\toprule
	&		&	KS$_1$	&	KS$_2$	&	KS$_3$	&	KS$_4$	&	KS$_5$	&	mean	&	sd	&	skewness	&	ex.kurtosis	&	rho	&	$\| var(\bar{g})^{-1/2}\bar{g}\|^2$	&	$\| \bar{g}\|^2$	\\
\midrule																											
Mnormal	&	MLE	&	0.079	&	0.077	&	0.079	&	0.074	&	0.080	&	0.000	&	0.000	&	0.698	&	36.632	&	0.000	&		&	1.07e-16	\\
\midrule																											
\multirow{3}{*}{MGH}	&	moments	&	0.056	&	0.096	&	0.073	&	0.077	&	0.081	&	0.000	&	0.000	&	55.958	&	364.456	&	0.001	&		&	1.73e-11	\\
	&	EM	&	0.027	&	0.014	&	0.022	&	0.018	&	0.021	&	0.000	&	0.003	&	4.970	&	193.856	&	0.014	&		&	5.97e-12	\\
	&	GMM	&	0.034	&	0.058	&	0.040	&	0.043	&	0.039	&	0.005	&	0.001	&	54.008	&	359.648	&	0.005	&	0.049	&	1.24e-6	\\
\midrule																											
\multirow{3}{*}{MNTS}	&	moments	&	0.061	&	0.045	&	0.067	&	0.060	&	0.054	&	0.001	&	0.000	&	3.032	&	33.685	&	0.030	&		&	1.55e-9	\\
	&	EM	&	0.028	&	0.015	&	0.022	&	0.019	&	0.022	&	0.000	&	0.005	&	0.639	&	36.708	&	0.015	&		&	3.43e-9	\\
	&	GMM	&	0.027	&	0.023	&	0.058	&	0.044	&	0.026	&	0.000	&	0.007	&	3.386	&	33.088	&	0.038	&	0.060	&	2.36e-9	\\
\midrule																											
\multirow{2}{*}{MMixedTS}	&	moments	&	0.035	&	0.017	&	0.032	&	0.011	&	0.039	&	0.000	&	0.000	&	0.020	&	4.613	&	0.781	&		&	5.69e-10	\\
	&	GMM	&	0.034	&	0.016	&	0.031	&	0.010	&	0.038	&	0.000	&	0.000	&	0.020	&	4.604	&	0.781	&	0.989	&	2.23e-9	\\
\midrule																											
\multirow{2}{*}{AlphaGH}	&	moments	&	0.096	&	0.096	&	0.102	&	0.100	&	0.100	&	0.000	&	0.001	&	8.318	&	13.481	&	0.021	&		&	2.07e-9	\\
	&	MLE + correlation	&	0.034	&	0.038	&	0.024	&	0.028	&	0.043	&	0.002	&	0.007	&	1.601	&	37.617	&	0.384	&		&	2.12e-7	\\
	&	GMM	&	0.079	&	0.074	&	0.068	&	0.060	&	0.086	&	0.002	&	0.004	&	8.483	&	13.233	&	0.028	&	0.074	&	1.98e-7	\\
\midrule																											
\multirow{2}{*}{RhoAlphaGH}	&	moments	&	0.154	&	0.110	&	0.121	&	0.103	&	0.207	&	0.001	&	0.031	&	0.001	&	39.408	&	0.103	&		&	1.91e-7	\\
	&	MLE + correlation	&	0.034	&	0.038	&	0.024	&	0.028	&	0.043	&	0.002	&	0.007	&	1.601	&	37.617	&	0.000	&		&	2.12e-7	\\
	&	GMM	&	0.088	&	0.061	&	0.063	&	0.047	&	0.104	&	0.001	&	0.006	&	0.280	&	39.873	&	0.065	&	0.104	&	2.12e-8	\\
\midrule																											
\multirow{2}{*}{MGVG}	&	moments	&	0.080	&	0.100	&	0.016	&	0.043	&	0.062	&	0.000	&	0.000	&	0.019	&	25.148	&	0.528	&		&	4.11e-7	\\
	&	MLE + correlation	&	0.014	&	0.016	&	0.013	&	0.012	&	0.018	&	0.000	&	0.003	&	0.417	&	54.392	&	1.035	&		&	5.83e-6	\\
	&	GMM	&	0.148	&	0.117	&	0.086	&	0.087	&	0.137	&	0.002	&	0.013	&	3.674	&	24.110	&	0.590	&	0.424	&	1.75e-7	\\
\midrule																											
\multirow{2}{*}{ICA LinearLG}	&	FastICA + moments	&	0.090	&	0.059	&	0.059	&	0.068	&	0.068	&	0.000	&	0.000	&	0.001	&	7.732	&	0.000	&		&	1.03e-9	\\
	&	FastICA + MLE	&	0.091	&	0.064	&	0.066	&	0.063	&	0.061	&	0.000	&	0.000	&	0.371	&	7.620	&	0.000	&		&	5.68e-10	\\
	&	FastICA + GMM	&	0.067	&	0.043	&	0.057	&	0.059	&	0.055	&	0.000	&	0.000	&	0.490	&	6.811	&	0.000	&	0.994	&	1.03e-9	\\
\midrule																											
\multirow{2}{*}{ICA LinearCTS}	&	FastICA + moments	&	0.091	&	0.066	&	0.061	&	0.067	&	0.079	&	0.000	&	0.000	&	0.000	&	7.741	&	0.000	&		&	8.24e-10	\\
	&	FastICA + MLE	&	0.102	&	0.077	&	0.068	&	0.077	&	0.090	&	0.000	&	0.000	&	0.146	&	5.942	&	0.000	&		&	9.57e-10	\\
	&	FastICA + GMM	&	0.091	&	0.066	&	0.061	&	0.067	&	0.079	&	0.000	&	0.000	&	0.000	&	7.741	&	0.000	&	0.998	&	8.24e-10	\\
\midrule																											
\multirow{2}{*}{PCA LinearCTS}	&	PCA + moments	&	0.034	&	0.035	&	0.046	&	0.036	&	0.044	&	0.000	&	0.031	&	0.579	&	56.988	&	0.066	&		&	1.10e-9	\\
	&	PCA + MLE	&	0.046	&	0.038	&	0.041	&	0.039	&	0.047	&	0.002	&	0.020	&	2.029	&	38.635	&	0.383	&		&	2.36e-7	\\
	&	PCA + GMM	&	0.032	&	0.033	&	0.045	&	0.034	&	0.041	&	0.000	&	0.029	&	0.577	&	60.167	&	0.068	&	0.844	&	1.22e-8	\\

\bottomrule
\end{tabular}
\caption[{\it Estimation errors}]{\label{tab:EstimationError}\footnotesize For each model and estimation approach we report the univariate Kolmogorov-Smirnov distance and the errors in fitting empirical moments in the period from July, 1 2003 to June, 29 2018. We estimate the model by considering the following algoritms: we minimize the distance between empirical and theoretical moments ({\it moments}), the expectation-maximization ({\it EM}), the generalized method of moments ({\it GMM}), the two-step procedure in which we first estimate the univariate margins by maximum likelihood estimation (MLE) and then we minimize the distance between empirical and theoretical correlations ({\it MLE + correlations}), the two-step procedure in which we first apply the FastICA or the PCA algoritm and then we minimize the errors between empirical and theoretical moments ({\it FastICA or PCA + moments}), the two-step procedure in which we first apply the FastICA or the PCA algoritm and then we estimate the standardized independent components by MLE or GMM ({\it FastICA or PCA + MLE or GMM}). While $\| var(\bar{g})^{-1/2}\bar{g}\|^2$ represents the GMM objective function at the optimal point, $\| \bar{g}\|^2$ represents the average distance between the empirical and the theoretical characteristic function with $q=50$.}
\end{scriptsize}
\end{center}
\end{sidewaystable}

\newpage

\newpage
\section*{Moments}

In this Appendix we provide the formulas of expected value, variance, skewness, excess kurtosis and correlation of the models described in the paper. Recall that the cumulant of order $j$ of a random variable $X$ with cumulant generating function $\varphi_{X}(u)$ can be computed as
$$
c_j(X) = \frac{\partial^j}{\partial u^j}\varphi_{X}(u)|_{u=0},
$$
and the following equalities hold
\begin{equation*}\label{ap:eq:moments}
\begin{split}
E[X] &= c_1(X),\\
var[X] &= c_2(X),\\
skew[X] &= \frac{c_3(X)}{c_2(X)^{3/2}},\\
kurt[X] &= 3 + \frac{c_4(X)}{c_2(X)^2}.\\
\end{split}
\end{equation*}

\subsection*{MNTS model}

\begin{equation*}
E\left[Y_{j,1}\right]=\mu_j +	E\left[S_{1}\right]\theta_j,
\end{equation*}

\begin{equation*}		
var\left[Y_{j,1}\right]=var\left[S_{1}\right]\left(\theta_j^2+ \frac{ \sigma_{j}^2 \lambda}{1-\omega}\right),
\end{equation*}

\begin{equation*}	
skew\left[Y_{j,1}\right]=skew\left[S_{1}\right]\left(\theta_j^3+ \frac{3\theta_j \sigma_{j}^2 \lambda}{2-\omega}\right)\left(\theta_j^2+ \frac{ \sigma_{j}^2 \lambda}{1-\omega}\right)^{-\frac{3}{2}},
\end{equation*}

\begin{equation*}		
kurt\left[Y_{j,1}\right]=3+\left(kurt\left[S_{1}\right]-3\right)\left[\theta_j^4+ \frac{3\sigma_{j}^2 \lambda}{3-\omega}\left(2 \theta_j^2+ \frac{\sigma_{j}^2 \lambda}{2-\omega}\right)\right]\left(\theta_j^2+ \frac{ \sigma_{j}^2 \lambda}{1-\omega}\right)^{-2},
\end{equation*}

\begin{equation*} 
corr\left[Y_{j,1}; Y_{k,1}\right]=\frac{\theta_j\theta_k+ \frac{ \sigma_{jk} \lambda}{1-\omega}}{\sqrt{\left(\theta_j^2+ \frac{\sigma_{j}^2 \lambda}{1-\omega}\right)\left(\theta_k^2+ \frac{ \sigma_{k}^2 \lambda}{1-\omega}\right)}},
\end{equation*}
where
\begin{equation*}
	E\left[S_{1}\right]=-\omega C\Gamma(-\omega) \lambda^{\omega -1},
\end{equation*}	
\begin{equation*}		
var\left[S_{1}\right]=\omega (\omega -1) C \Gamma(-\omega) \lambda^{\omega -2},
\end{equation*}
\begin{equation*}	
skew\left[S_{1}\right]=(2-\omega) \left[\omega (\omega -1) C \Gamma(-\omega) \lambda^{\omega}\right]^{-\frac{1}{2}},
\end{equation*}
\begin{equation*}		
kurt\left[S_{1}\right]=3+(\omega -2)(\omega -3)\left[\omega (\omega -1) C \Gamma(-\omega) \lambda^{\omega}\right]^{-1}.
\end{equation*}

\subsection*{MGH model}

\begin{equation*}
E\left[Y_{j,1}\right]=\mu_j + E\left[G_{1}\right]\theta_{j},
\end{equation*}
\begin{equation*}		
var\left[Y_{j,1}\right]=E\left[G_{1}\right]\sigma_j^2+var\left[G_{1}\right] \theta_j^2,
\end{equation*}	
\begin{equation*}		
skew\left[Y_{j,1}\right]=\frac{c_3\left[Y_{j,1}\right]}{var\left[Y_{j,1}\right]^{3/2}},
\end{equation*}	
\begin{equation*}		
kurt\left[Y_{j,1}\right]=3+\frac{c_4\left[Y_{j,1}\right]}{var\left[Y_{j,1}\right]^{2}},
\end{equation*}
\begin{equation*}	
corr\left[Y_{j,1}; Y_{k,1}\right]=\frac{\sigma_{jk}+ \theta_j\theta_k\Delta\left(\frac{\chi}{\psi}\right)^\frac{1}{2}}{\sqrt{\left[\sigma_{j}^2+ \theta_{j}^2\Delta\left(\frac{\chi}{\psi}\right)^\frac{1}{2}\right]\left[\sigma_{k}^2+ \theta_{k}^2\Delta\left(\frac{\chi}{\psi}\right)^\frac{1}{2}\right]}},
\end{equation*}
where
\begin{equation*}	
c_3\left[Y_{j,1}\right]=3 var\left[G_{1}\right]\theta_j\sigma_j^2+  c_3\left[G_{1}\right]\theta_j^3,
\end{equation*}
\begin{equation*}		
c_4\left[Y_{j,1}\right]=3 var\left[G_{1}\right]\sigma_j^4+ 6c_3\left[G_{1}\right]\theta_j^2\sigma_j^2+c_4\left[G_{1}\right]\theta^4_j,
\end{equation*}
\begin{equation*}		
cov\left[Y_{j,1}; Y_{k,1}\right]=E\left[G_{1}\right]\sigma_{jk}+var\left[G_{1}\right] \theta_j\theta_k,
\end{equation*}
\begin{equation*}
	\Delta=\left(\frac{K_{\epsilon +2}\left(\sqrt{\chi\psi}\right)}{K_{\epsilon +1}\left(\sqrt{\chi\psi}\right)}-\frac{K_{\epsilon +1}\left(\sqrt{\chi\psi}\right)}{K_{\epsilon}\left(\sqrt{\chi\psi}\right)}\right),
\end{equation*}
and with
\begin{equation*}
E\left[G_{1}\right]=\left(\frac{\chi}{\psi}\right)^\frac{1}{2}\frac{ K_{\epsilon+1}\left(\sqrt{\chi\psi}\right) }{ K_{\epsilon}\left(\sqrt{\chi\psi}\right)},
\end{equation*}
\begin{equation*}		
var\left[G_{1}\right]=\left(\frac{\chi}{\psi}\right)\left[\frac{K_{\epsilon+2}\left(\sqrt{\chi\psi}\right)}{K_{\epsilon}\left(\sqrt{\chi\psi}\right)}-\left(\frac{K_{\epsilon+1}\left(\sqrt{\chi\psi}\right)}{K_{\epsilon}\left(\sqrt{\chi\psi}\right)}\right)^2\right],
\end{equation*}
\begin{equation*}
c_3\left[G_{1}\right] =\left(\frac{\chi}{\psi}\right)^\frac{3}{2}\left[\frac{K_{\epsilon+3}\left(\sqrt{\chi\psi}\right)}{K_{\epsilon}\left(\sqrt{\chi\psi}\right)}-\frac{3K_{\epsilon+2}\left(\sqrt{\chi\psi}\right)K_{\epsilon+1}\left(\sqrt{\chi\psi}\right)}{K^2_{\epsilon}\left(\sqrt{\chi\psi}\right)}+2\left(\frac{K_{\epsilon+1}\left(\sqrt{\chi\psi}\right)}{K_{\epsilon}\left(\sqrt{\chi\psi}\right)}\right)^3\right],
\end{equation*}	
\begin{equation*}
\begin{split}		
c_4\left[G_{1}\right]&= \left(\frac{\chi}{\psi}\right)^2\left[\frac{K_{\epsilon+4}\left(\sqrt{\chi\psi}\right)}{K_{\epsilon}\left(\sqrt{\chi\psi}\right)}-\frac{4K_{\epsilon+3}\left(\sqrt{\chi\psi}\right)K_{\epsilon+1}\left(\sqrt{\chi\psi}\right)}{K^2_{\epsilon}\left(\sqrt{\chi\psi}\right)}-3\left(\frac{K_{\epsilon+2}\left(\sqrt{\chi\psi}\right)}{K_{\epsilon}\left(\sqrt{\chi\psi}\right)}\right)^2 \right]+\\&+6\left(\frac{\chi}{\psi}\right)^2\left[\frac{2 K_{\epsilon+2}\left(\sqrt{\chi\psi}\right)K^2_{\epsilon+1}\left(\sqrt{\chi\psi}\right)}{K^3_{\epsilon}\left(\sqrt{\chi\psi}\right)}-\left(\frac{K_{\epsilon+1}\left(\sqrt{\chi\psi}\right)}{K_{\epsilon}\left(\sqrt{\chi\psi}\right)}\right)^4 \right].
\end{split}
\end{equation*}

\subsection*{\texorpdfstring{$\alpha$}{$\alpha$}GH model}

The margins are GH distributed with parameters $(\mu_j, \theta_j, \sigma_j, \chi_j, \psi_j, \epsilon)$, for each $j=1,..,n$. While marginal moments are as in MGH case, the correlations are given by
\begin{equation*}	
corr\left[Y_{j,1}; Y_{k,1}\right]=\frac{4a\left(\psi_j\psi_k \right)^{-1}\theta_j\theta_k} {\sqrt{\left[E\left[G_{1,j}\right]\sigma_j^2+var\left[G_{1,j}\right] \theta_j^2\right]\left[E\left[G_{1,k}\right]\sigma_k^2+var\left[G_{1,k}\right] \theta_k^2\right]}},
\end{equation*}
where
\begin{equation} \label{ap:eq:expectGIG}
E\left[G_{j,1}\right]=\left(\frac{\chi_j}{\psi_j}\right)^\frac{1}{2}\frac{ K_{\epsilon+1}\left(\sqrt{\chi_j\psi_j}\right) }{ K_{\epsilon}\left(\sqrt{\chi_j\psi_j}\right)},
\end{equation}
and
\begin{equation} \label{ap:eq:varGIG}		
var\left[G_{j,1}\right]=\left(\frac{\chi_j}{\psi_j}\right)\left[\frac{K_{\epsilon+2}\left(\sqrt{\chi_j\psi_j}\right)}{K_{\epsilon}\left(\sqrt{\chi_j\psi_j}\right)}-\left(\frac{K_{\epsilon+1}\left(\sqrt{\chi_j\psi_j}\right)}{K_{\epsilon}\left(\sqrt{\chi_j\psi_j}\right)}\right)^2\right].
\end{equation}

\subsection*{\texorpdfstring{$\rho\alpha$}{$\rho\alpha$}GH model}

The margins are GH distributed with parameters $(\mu_j, \theta_j, \sigma_j, \chi_j, \psi_j, \epsilon)$, for each $j=1,..,n$. While marginal moments are as in MGH case, the correlations are given by
\begin{equation*}
corr\left[Y_{j,1}; Y_{k,1}\right]=\frac{2a\left(\psi_j\psi_k \right)^{-1} \left(\sigma_j\sigma_k\rho_{jk}\sqrt{\psi_j}\sqrt{\psi_k}+2\theta_j\theta_k\right)} {\sqrt{\left[E\left[G_{1,j}\right]\sigma_j^2+var\left[G_{1,j}\right] \theta_j^2\right]\left[E\left[G_{1,k}\right]\sigma_k^2+var\left[G_{1,k}\right] \theta_k^2\right]}}
\end{equation*}
where $E\left[G_{j,1}\right]$ and $var\left[G_{j,1}\right]$ are as in equations (\ref{ap:eq:expectGIG}) and (\ref{ap:eq:varGIG}).

\subsection*{MMixedTS model}

\begin{equation*}
E\left[Y_{j,1}\right]=\mu_{j}+\beta_{j}\frac{c_{j}+\bar{n}}{m_{j}},
\end{equation*}

\begin{equation*}
var\left[Y_{j,1}\right]=\left(1+\frac{\beta_{j}^{2}}{m_{j}}\right)\frac{\left(c_{j}+\bar{n}\right)}{m_{j}},
\end{equation*}

\begin{equation*}		
skew\left[Y_{j,1}\right]=\frac{c_3\left[Y_{j,1}\right]}{var\left[Y_{j,1}\right]^{3/2}},
\end{equation*}	
\begin{equation*}		
kurt\left[Y_{j,1}\right]=3+\frac{c_4\left[Y_{j,1}\right]}{var\left[Y_{j,1}\right]^{2}},
\end{equation*}
\begin{equation*}
corr\left[Y_{j,1},Y_{k,1}\right]=\frac{\frac{\beta_{j}\beta_{k}}{m_{j}m_{k}}\bar{n}}{\sqrt{\left(1+\frac{\beta_{j}^{2}}{m_{j}}\right)\frac{\left(c_{j}+\bar{n}\right)}{m_{j}}}\sqrt{\left(1+\frac{\beta_{k}^{2}}{m_{k}}\right)\frac{\left(c_{k}+\bar{n}\right)}{m_{k}}}} .
\end{equation*}
where
\begin{equation*}
c_{3}\left[Y_{j,1}\right] =\left[\left(2-\alpha_{j}\right)\frac{\lambda_{+,j}^{\alpha_{j}-3}-\lambda_{-,j}^{\alpha_{j}-3}}{\lambda_{+,j}^{\alpha_{j}-2}+\lambda_{-,j}^{\alpha_{j}-2}}+\left(3+2\frac{\beta_{j}^{2}}{m_{j}}\right)\frac{\beta_{j}}{m_{j}}\right]\frac{\left(l_{j}+\bar{n}\right)}{m_{j}},
\end{equation*}
\begin{align*}
c_{4}\left[Y_{j,1}\right] & =\beta_{j}^{4}\left(3+\frac{6}{l_{j}+\bar{n}}\right)\frac{\left(l_{j}+\bar{n}\right)^{2}}{m_{j}^{4}}+6\beta_{j}^{2}\frac{l_{j}+\bar{n}}{m_{j}^{3}}\left(l_{j}+\bar{n}+2\right)+\nonumber \\
 & +4\beta_{j}\left(2-\alpha_{j}\right)\left(\frac{\lambda_{+,j}^{\alpha_{j}-3}-\lambda_{-,j}^{\alpha_{j}-3}}{\lambda_{+,j}^{\alpha_{j}-2}+\lambda_{-,j}^{\alpha_{j}-2}}\right)\frac{l_{j}+\bar{n}}{m_{j}^{2}}+\left(3-\alpha_{j}\right)\left(2-\alpha_{j}\right)\left(\frac{\lambda_{+,j}^{\alpha_{j}-4}+\lambda_{-,j}^{\alpha_{j}-4}}{\lambda_{+,j}^{\alpha_{j}-2}+\lambda_{-,j}^{\alpha_{j}-2}}\right)\frac{l_{j}+\bar{n}}{m_{j}}.
\end{align*}

\subsection*{MGVG model}

\begin{equation*}
E\left[Y_{j,1}\right]=\mu_j + E\left[\hat G_{j,1}\right]\theta_{j}=\mu_j +\theta_{j},
\end{equation*}
\begin{equation*}		
var\left[Y_{j,1}\right]=E\left[\hat G_{j,1}\right]\sigma_j^2+var\left[\hat G_{j,1}\right] \theta_j^2=\sigma_j^2+\left(1-k_j\right)q_j\theta_j^2+2k_jp_j\theta_j^2,
\end{equation*}	
\begin{equation*}		
skew\left[Y_{j,1}\right]=\frac{c_3\left[Y_{j,1}\right]}{var\left[Y_{j,1}\right]^{3/2}},
\end{equation*}	
\begin{equation*}		
kurt\left[Y_{j,1}\right]=3+\frac{c_4\left[Y_{j,1}\right]}{var\left[Y_{j,1}\right]^{2}},
\end{equation*}
\begin{equation*}	
corr\left[Y_{j,1}; Y_{k,1}\right]=\frac{\sigma_j\sigma_k\rho_{jk}\left(c_1\sqrt{q_jq_k}+c_2\sqrt{p_jp_k}\right)+\theta_j\theta_k \left(c_1q_jq_k+2c_2p_jp_k\right)} {\sqrt{var\left[Y_{j,1}\right]var\left[Y_{k,1}\right]}}
\end{equation*}
where
\begin{equation*}
\begin{split}	
c_3\left[Y_{j,1}\right]&=3 var\left[\hat G_{j,1}\right]\theta_j\sigma_j^2+ c_3\left[\hat G_{j,1}\right]\theta_j^3 \\
\end{split}
\end{equation*}
\begin{equation*}
\begin{split}		
c_4\left[Y_{j,1}\right]&=3 var\left[\hat G_{j,1}\right]\sigma_j^4+ 6c_3\left[\hat G_{j,1}\right]\theta_j^2\sigma_j^2+c_4\left[\hat G_{j,1}\right]\theta^4_j\\
\end{split}
\end{equation*}
\begin{equation*}		
cov\left[Y_{j,1}; Y_{k,1}\right]=E\left[\hat G_{j,1}\right]\sigma_{jk}+var\left[\hat G_{j,1}\right] \theta_j\theta_k,
\end{equation*}
and with
\begin{equation*}
E\left[ \hat G_{j,1}\right]=1,
\end{equation*}	
\begin{equation*}		
var\left[\hat G_{j,1}\right]=\left(1-k_j\right)q_j+2k_jp_j,
\end{equation*}
\begin{equation*}		
c_3\left[\hat G_{j,1}\right]=2(1-k_j)q_j^2+6k_jp_j^2,
\end{equation*}
\begin{equation*}		
c_4\left[\hat G_{j,1}\right]=6\left(1-k_j\right)q_j^3+24k_jp_j^3.
\end{equation*}

\end{document}